\documentclass[11pt]{article}
\usepackage{amsmath}
\usepackage{amssymb}
\usepackage{amsfonts}
\usepackage{amsthm}   
\usepackage{color}
\usepackage{bm}        
\usepackage{enumerate}
\usepackage{geometry}
\newtheorem{theorem}{Theorem}

\newtheorem{condition}{Condition}
\newtheorem{proposition}{Proposition}
\newtheorem{remark}{Remark}
\usepackage{graphicx,psfrag,epsf}
\usepackage{indentfirst}
\usepackage{booktabs,multicol,multirow,rotating}
\usepackage{natbib}
\usepackage{hyperref}

\newtheorem{corollary}{Corollary}

\usepackage{xr}

\title{Adaptive Two-Sample Testing for Regression Discrepancy and Model Misspecification in High Dimensions}

\author{Yu Peng$^1$, Jiaqi Huang$^2$, Falong Tan$^3$, and Lixing Zhu$^1$\footnote{Lixing Zhu and Falong Tan are co-corresponding authors (lzhu@bnu.edu.cn; falongtan@hnu.edu.cn). Yu Peng and Jiaqi Huang are co-first authors. The authors are grateful to Dr. Chuhan Wang for valuable consultations regarding the theoretical investigation. The research was supported by the  grant  (NSFC12471276) from the National Natural Science Foundation of China  and the grant (CI2023C063YLL) from the Scientific and Technological Innovation Project of China Academy of Chinese Medical Science.}\\
$^1$ Department of Statistics, Beijing Normal University at Zhuhai, China\\
$^2$ Department of Statistics, University of Michigan, USA \\
$^3$ Department of Statistics and Data Science, Hunan University, China}
\date{\vspace{-5ex}}

\begin{document}

\maketitle

\begin{abstract}
In two-sample regression comparisons, testing for equality is often too restrictive because minor discrepancies in functional form are unavoidable in practice. We study whether the functional discrepancy falls within a practically acceptable tolerance. Such tolerance-based comparison can, however, be unreliable when either regression model is misspecified. To address both issues, we propose a unified adaptive-to-model testing framework for high-dimensional regressions that integrates model checking with functional similarity testing for the two populations.
For functional similarity, we develop an adaptive boundary-proxy test whose rejection probability converges to the nominal level at every fixed configuration of the composite null hypothesis, overcoming the conservatism of conventional least favorable configuration tests. For model misspecification, we use projection-based tests for sparse high-dimensional regression models. These components are combined through adaptive gating into a joint test. We investigate its asymptotic null distribution and power properties under global and local alternatives. Numerical studies examine its finite-sample performance.
\end{abstract}

\section{Introduction}
Given two independent populations, a central question is whether some of the underlying distributional or structural characteristics are identical, or sufficiently close to each other. Classical two-sample procedures typically focus on comparisons of means, covariance matrices, distributions, or other finite-dimensional parameters. In many modern applications, however, the object of interest is a regression relationship, and the goal is to determine whether the relationships between a response and a set of covariates are sufficiently similar across two populations. Such problems arise naturally in many areas, including transfer learning with one source dataset, where one seeks to determine whether information from a source population can be appropriately used for a target population.

For parametric regression models, a natural way to compare two populations is through the discrepancy between their regression functions. Rather than comparing parameter vectors directly, one may quantify the difference between the corresponding regression functions using a functional distance, such as the squared $L_2$-distance. Distance-based comparisons of regression functions have been extensively studied in the literature; see, among others, \cite{munk1998nonparametric, dette2001nonparametric, dette2018equivalence, mollenhoff2020equivalence}. Depending on the scientific objective, the distance may be used to formulate either an exact comparison or a threshold-based testing problem. The latter formulation is particularly relevant to the present paper. Instead of requiring the functional distance to be exactly zero, a prespecified threshold can be introduced to distinguish practically negligible differences from relevant departures. In particular, \citet{dette2020testing} considered two-sample functional problems in which an $L_2$-distance is compared with a relevance threshold. Motivated by this perspective, we consider a tolerance testing problem for two regression functions, where the null hypothesis allows a certain degree of discrepancy as long as it remains within a prespecified acceptable range.

Distance-based similarity testing, however, addresses only one aspect of the two-sample regression problem. Its validity depends on the adequacy of the regression models used to construct the functional distance. If the working regression model is misspecified in either population, the estimated discrepancy may partly reflect model approximation error rather than a genuine difference between the underlying regression relationships. As a result, a comparison based solely on the functional distance may lead to misleading conclusions. This issue is particularly relevant in transfer learning, where a small estimated source-target discrepancy is not sufficient to support information sharing if the regression model in either domain is itself misspecified.

To this end, we extend the distance-based inequality testing problem by simultaneously incorporating model misspecification assessment for the two populations. The overall testing problem requires not only that the functional discrepancy remain below the prespecified tolerance level, but also that the working regression models provide adequate representations of the corresponding conditional mean functions. To achieve this goal, we construct an adaptive-to-model testing procedure that combines the adaptive similarity statistic with model-checking statistics for the two samples through a gating mechanism. Under correct model specification, the model-checking components are asymptotically inactive, whereas under model misspecification the corresponding components can be adaptively activated. The resulting procedure is therefore able to respond to excessive functional discrepancy, model misspecification in either population, or combinations of these departures. Moreover, the proposed framework is developed for sparse high-dimensional regression settings, allowing the predictor dimension to be comparable to or substantially larger than the sample sizes.

Transfer learning may be one of the potential application scenarios of this joint testing framework. In a single-source transfer problem, the two samples naturally correspond  to a target domain and a source domain. Successful transfer generally requires a certain degree of compatibility between the two domains, whereas substantial heterogeneity can lead to negative transfer \citep{pan2009survey,weiss2016survey,li2022transfer,tian2023transfer}. Existing high-dimensional transfer learning methods typically impose some notion of source-target similarity when deciding whether source information should be incorporated into the target analysis. From the perspective of the present paper, such compatibility can be examined through a two-sample regression test: the regression functions should be sufficiently close, while the working models used to describe them should also be correctly specified. 

The main contributions of this paper are threefold. First, we develop an adaptive boundary-proxy test for distance-based tolerance testing that alleviates the over-conservatism of the conventional inequality testing procedure based on the least favorable configuration (LFC) in the interior of the null region while retaining its favorable locally asymptotically uniformly most powerful (LAUMP) property near the boundary. Second, we go beyond functional similarity testing by explicitly incorporating model misspecification assessment into the two-sample comparison, thereby yielding an adaptive-to-model joint testing framework. Third, the proposed methodology is developed for sparse high-dimensional regression models, allowing both functional similarity and model adequacy to be assessed when the predictor dimension may substantially exceed the sample size. We establish the asymptotic properties of the proposed test statistics and investigate their power under various global and local alternatives.

The remainder of this paper is organized as follows. Section 2 introduces the problem setup and formulates the joint hypotheses for functional similarity and model specification. Section 3 develops the adaptive boundary-proxy test for functional similarity and establishes its asymptotic properties. Section 4 constructs the adaptive-to-model joint test, introduces the model specification components, and analyzes the behavior of the proposed statistic under various global and local alternatives. Section 5 discusses the application to transfer learning. Section 6 presents the simulation studies and the real-data application, and Section 7 discusses extensions to multi-source transfer learning and other directions for future research. All technical proofs are provided in the supplementary material.

\noindent {\bf Notation.} Bold capital letters (e.g., $\mathbf{A}$) represent matrices, and for a matrix $\mathbf{A}$ we write $\|\mathbf{A}\|$ for its Frobenius norm. For a vector $v$ and $q \geq 1$, $\|v\|_q$ denotes its $\ell_q$-norm and $\|v\|_0$ denotes the number of its nonzero components. Let $\mathcal{S}^{p-1}=\{\alpha\in\mathbb{R}^p:\|\alpha\|_2=1\}$ denote the unit sphere in $\mathbb{R}^p$.
We use $\mathbb{I}(\cdot)$ to denote the indicator function and $\mathbb{P}(\cdot)$ the probability measure. 
For any function $g(x)$, let $\dot{g}(x) = \frac{dg(x)}{dx}$ and $\ddot{g}(x) =  \frac{d\dot{g}(x)}{dx}$.
We write $ \stackrel{P}{\longrightarrow} $ and $ \stackrel{D}{\longrightarrow} $ to denote convergence in probability and  convergence in distribution, respectively. The notation $X \sim F$ indicates that $X$ follows the distribution $F$. We use $s \ll p$ to denote $s/p \to 0$. Moreover, $a \vee b= \max(a,b)$, $a \wedge b= \min(a,b)$, $a \vee b \vee c = \max\{a, b, c\}$, and $a \wedge b \wedge c = \min\{a, b, c\}$.
Throughout this paper, $C$ represents a generic constant independent of the sample size $n$, and its value may change across appearances.

\section{Problem Setup and Preliminary Definitions}
In this paper, we consider a two-sample regression problem. Suppose that we have two independent datasets  $ \{(X^{(0)}_i,Y^{(0)}_i)\}_{i=1}^{n_0} $
and $ \{(X^{(1)}_i,Y^{(1)}_i)\}_{i=1}^{n_1}$ , where $(X_i^{(l)}, Y_i^{(l)}) \in \mathbb{R}^{p+1}$ for $l=0, 1$ are independent and identically distributed within each dataset. The underlying regression model is
$$ Y^{(l)} = m_l(X^{(l)})+\epsilon^{(l)}, \quad l=0, 1, $$ 
where $m_l(x) = \mathbb{E}(Y^{(l)}|X^{(l)}=x)$ denotes the unknown regression function and $\epsilon^{(l)}$ denotes the random error with zero mean and finite variance. Given a parametric family of functions $\{g_l(\cdot,\theta^{(l)}): \theta^{(l)} \in \Theta_l\subset\mathbb{R}^{q_l}\}$, our objective is two-fold. The first is model checking, which is used to determine whether each parametric model $g_l(X^{(l)},\theta^{(l)})$ accurately represents its respective data-generating process. 
The other is similarity evaluation. The issue is to assess whether the discrepancies between models are negligible. If the models are sufficiently similar, the data from different groups may be pooled. This also serves as the core motivation for transfer learning, where structural commonalities allow knowledge to be adapted from source to target domains. When applied to single-source transfer learning, the two populations $0$ and $1$ can be interpreted as the target and source domains, respectively.


We now formulate the hypothesis testing problem to simultaneously assess the model adequacy and similarity between two regression models in high-dimensional settings, where the predictor dimension may substantially exceed the sample size. 
Recall that for the model checking problem, our objective is to test whether the mean function $m_l(\cdot)$ belongs to a given parametric class of functions $\{g_l(\cdot, \theta^{(l)}): \theta^{(l)} \in \Theta_l \subset \mathbb{R}^{q_l}\}$ for $l=0, 1$. 
In this paper, we focus on testing for model misspecification for generalized linear models (GLMs). Extensions to more general model checking problems are straightforward.
The null and alternative hypotheses for model checking for the populations $0$ and $1$ are
\begin{align}\label{source_hypo}
\begin{split}
& H_0^{(0)}: \mathbb{P} \{ \mathbb{E}(Y^{(0)}|X^{(0)}) 
= g_0( (\theta_0^{(0)})^{\top} X^{(0)}) \} = 1, \ \text{for some} \ \theta_0^{(0)} \in \Theta_0, \quad \text{versus} \\
& H_1^{(0)}: \mathbb{P} \{ \mathbb{E}(Y^{(0)}|X^{(0)}) \neq g_0( \theta^{\top} X^{(0)}) \} > 0, \ \text{for any} \ \theta \in \Theta_0,
\end{split}
\end{align}
and 
\begin{align}\label{target_hypo}
\begin{split}
& H_0^{(1)}: \mathbb{P} \{ \mathbb{E}(Y^{(1)}|X^{(1)}) = g_1((\theta_0^{(1)})^{\top} X^{(1)}) \} = 1, \ \text{for some} \ \theta_0^{(1)} \in \Theta_1,  \quad
\text{versus} \\
& H_1^{(1)}: \mathbb{P} \{ \mathbb{E}(Y^{(1)}|X^{(1)}) \neq g_1(\theta^{\top} X^{(1)}) \} > 0, \ \text{for any} \ \theta \in \Theta_1,
\end{split}
\end{align}
respectively, where $\Theta_l \subset \mathbb{R}^p $ is a compact set and $g_l(\cdot)$ is a known inverse link function. To ensure the model checking problem is well-defined in high-dimensional settings when the predictor dimension is larger than the sample size, we follow the framework from \cite{Jankova2020} and \cite{tan2025b} and impose sparsity under both the null and alternatives. Therefore, under the null $H_0^{(l)}$, we assume $\| \theta_0^{(l)} \|_0 = s_l \ll p$ for $l=0, 1$. 
Let $S^{(l)} \subset \{1, \cdots, p \}$ be the active set, i.e., the set of indices of the components of $X^{(l)}$ truly related to the response $Y^{(l)}$.
Under $H_0^{(l)}$, the true regression parameter $\theta_0^{(l)}$ is sparse, and the active set becomes $S^{(l)}=\{1 \leq j \leq p: \theta_{0j}^{(l)} \neq 0 \}$, where $\theta_{0j}^{(l)}$ denotes the $j$-th component of $\theta_0^{(l)}$.

The parameters $\theta_0^{(0)}$ and $\theta_0^{(1)}$ are unknown, and we estimate them under the high-dimensional GLM setting.
For $l=0,1$, let $\hat{\theta}^{(l)}$ be a penalized estimator of $\theta_0^{(l)}$ under the GLM setting, that is,
\begin{eqnarray}\label{penal-estimator}
\hat{\theta}^{(l)} 
= \mathop{\arg\min}\limits_{\theta \in \Theta_l} \Big\{ L_n^{(l)}(\theta) + \sum_{j=1}^p p_{\lambda}(|\theta_j|) \Big\},
\end{eqnarray}
where $L_n^{(l)}(\theta) = n_l^{-1} \sum_{i=1}^{n_l} \rho(Y_i^{(l)}, \theta^{\top}X_i^{(l)})$ with a general loss function $\rho(\cdot, \cdot)$, $p_{\lambda}(\cdot)$ is a penalty function, and $\lambda \geq 0$ is the regularization parameter.
Depending on the penalty function, the minimizer in (\ref{penal-estimator}) yields many popular sparse estimators, such as the GLM Lasso estimator, the GLM SCAD estimator, and the GLM MCP estimator. 
Under the alternative $H_1^{(l)}$, the GLMs are misspecified. Therefore, to investigate the asymptotic properties of $\hat{\theta}^{(l)}$ and the corresponding test statistics under the alternatives (misspecified GLMs), we define the pseudo-true parameters:
\begin{align}\label{tilde_theta}
\tilde{\theta}^{(l)} = \mathop{\arg\min}_{\theta \in \Theta_l} \mathbb{E}[\rho(Y^{(l)}, \theta^{\top}X^{(l)})],
\end{align} 
where $l=0,1$. Let $\tilde{S}^{(l)}=\{1 \leq j \leq p: \tilde{\theta}_{j}^{(l)} \neq 0 \}$ be the support of $\tilde{\theta}^{(l)}$ and $\tilde{s}^{(l)} = |\tilde{S}^{(l)}|$, where $\tilde{\theta}_{j}^{(l)}$ denotes the $j$-th component of $\tilde{\theta}^{(l)}$.

\begin{remark}
Under the null hypothesis $H_0^{(l)}$ and certain regularity conditions, the pseudo-true parameter $\tilde{\theta}^{(l)}$ generally coincides with the true parameter $\theta_0^{(l)}$, that is, $\tilde{\theta}^{(l)} = \theta_0^{(l)}$ for $l=0, 1$. Under the alternative hypothesis $H_1^{(l)}$, $\tilde{\theta}^{(l)}$ typically depends on the cumulative distribution functions of $Y^{(l)}$ and $X^{(l)}$. 
\citet{Buhlmann2015} investigated the asymptotic properties of the lasso estimator $\hat{\theta}^{(l)}$ under high-dimensional misspecified linear models. They showed that the support $\tilde{S}^{(l)} $ of $\tilde{\theta}^{(l)}$ satisfies $\tilde{S}^{(l)} \subset S^{(l)}$ when the predictor $X^{(l)}$ follows a Gaussian distribution with a positive definite covariance matrix.
\cite{lu2012} established the oracle property of the adaptive lasso estimator $\hat{\theta}^{(l)}$ under the misspecified GLM when the predictor dimension $p$ is fixed. They showed that $\tilde{S}^{(l)} = S^{(l)}$ if the true model is also a GLM with a misspecified link function and the linearity condition is satisfied for the predictor $X^{(l)}$, that is, $E(\theta^{\top}X^{(l)}|\theta_0^{\top}X^{(l)} = \theta_0^{\top}x) $ exists and is linear in $\theta_0^{\top}x$ for all $\theta \in \mathbb{R}^p$.
\end{remark}

For functional similarity, a conventional formulation is to test the exact
equality of two regression functions.
However, in practical applications, requiring exact equality of two regression functions is often unnecessarily restrictive and statistically unrealistic. By contrast, we propose an inequality-based testing framework which is more flexible and allows for minor and acceptable differences between the regression models. 
Specifically, let $(\mathcal{M},\varrho)$ be a semi-metric space of real-valued functions $g:\mathbb{R}^p \xrightarrow{} \mathbb{R}$ equipped with a semi-metric $\varrho$. We assume $g_l(x^{\top} \theta) \in \mathcal{M}$ for all $\theta \in \Theta_l$ and $l=0, 1$. In this paper, we are particularly interested in the semi-metric space of all continuous functions with $L_2$-distances
$$
\varrho\{ g_0(x^{\top} \theta^{(0)}), g_1(x^{\top} \theta^{(1)}) \} = \int_{\mathcal{X}} |g_0(x^{\top} \theta^{(0)})-g_1(x^{\top} \theta^{(1)})|^2 dF(x),
$$
where $F(x) = \zeta F_0(x)+ (1-\zeta) F_1(x) $ is a weighted mixture distribution of $X^{(0)}$ and $X^{(1)}$, $F_l(x)$ is the cumulative distribution function of $X^{(l)}$, and $\zeta \in (0, 1)$ is the limit of the ratio $n_0/(n_0+n_1)$, and $\mathcal{X}$ is the common compact support of $F_0$ and $F_1$. For notational simplicity, we write 
\begin{align}\label{d}
d = \int_{\mathcal{X}} |g_0(x^{\top} \tilde{\theta}^{(0)})-g_1(x^{\top} \tilde{\theta}^{(1)})|^2 dF(x), 
\end{align}
where $\tilde{\theta}^{(l)}$ is the pseudo-true parameter defined in (\ref{tilde_theta}).
Therefore, the null and alternative hypotheses for the similarity of the regression functions of two populations are 
\begin{align}\label{inqu-hypo}
 H_0^{(\Delta)}: 0< d \leq \Delta, \quad \quad \text{versus} \quad \quad
 H_1^{(\Delta)}: d > \Delta,
\end{align}
where $\Delta > 0$ is a tolerance threshold determined by the specific practical requirement of the application.

We now combine the model checking and functional similarity components into a joint testing problem. We restrict our attention to data-generating processes for which $d>0$. Within this restricted class, the joint hypotheses are
\begin{align}\label{H0}
\begin{split}
H_0 : H_0^{(0)}\cap H_0^{(1)}\cap H_0^{(\Delta)},   \quad \text{versus}\quad
H_1 : H_1^{(0)}\cup H_1^{(1)}\cup H_1^{(\Delta)}.
\end{split}
\end{align}
Thus, the null hypothesis $H_0$ holds if and only if the component null hypotheses $H_{0}^{(0)}$, $H_{0}^{(1)}$, and $ H_{0}^{(\Delta)} $ hold simultaneously, whereas the alternative hypothesis $H_1$ corresponds to the case in which at least one of $H_{1}^{(0)}$, $H_{1}^{(1)} $, and $H_{1}^{(\Delta)}$ holds.
Recall the discrepancy measure $d = \int_{\mathcal{X}} |g_0(x^{\top} \tilde{\theta}^{(0)})-g_1(x^{\top} \tilde{\theta}^{(1)})|^2 dF(x)$, which is defined in terms of the pseudo-true parameters $\tilde{\theta}^{(l)}$. Under the joint null hypothesis $H_0^{(0)} \cap H_0^{(1)}$, this quantity reduces to the $L_2$-distance between the two regression functions, $g_0(x^{\top} \theta_0^{(0)})$ and $g_1(x^{\top}\theta_0^{(1)})$. Conversely, if a component null hypothesis $H_0^{(l)}$ is false, the true parameter $\theta_0^{(l)}$ does not exist due to model misspecification. In such cases, $\tilde{\theta}^{(l)}$ represents the best-fitting parameter for the regression function $g_l(x^{\top} \theta^{(l)})$ within the specified class, thereby ensuring $d$ remains well defined.

\begin{remark}\label{d=0}
We explicitly exclude the case $d=0$ from the joint hypotheses (\ref{H0}). If $d=0$ and $g_l(\cdot)$ is smooth for $l=0, 1$, then $g_0(x^{\top} \tilde{\theta}^{(0)}) = g_1(x^{\top} \tilde{\theta}^{(1)})$ almost surely and the problem reduces to the case considered in \cite{wang2026model}. As detailed in the test construction, the asymptotic normality of $\hat d$ (to be defined later) is valid only when $d>0$. In the degenerate case $d=0$, the limiting distribution depends on the asymptotic properties of the estimated parameters $\hat \theta^{(0)}$ and $\hat \theta^{(1)}$, resulting in a convergence rate faster than $1/\sqrt{n}$; see Remark~{S1} in the Supplementary Material. Consequently, the components of our final tests (defined in Sections 3 and 4) related to $d$ become asymptotically negligible in this scenario. 
\end{remark}
\vspace{-8pt}

\section{Similarity Testing}\label{sec:similarity}
\subsection{Discrepancy estimation}
We first construct an estimator of the functional discrepancy and
establish its asymptotic properties.
Recall that the null and alternative hypotheses for functional similarity are 
\begin{align*}
 H_0^{(\Delta)}: 0< d \leq \Delta \quad \quad \text{versus} \quad \quad 
 H_1^{(\Delta)}: d > \Delta,
\end{align*}
where $d = \int_{\mathcal{X}} |g_0(x^{\top} \tilde{\theta}^{(0)})- g_1(x^{\top} \tilde{\theta}^{(1)})|^2dF(x) $. Throughout the paper, we restrict attention to fixed interior points of $H_0^{(\Delta)}$ and do not consider local sequences of $d$ drifting toward the boundary of the null region.
By substituting the population-specific estimator $\hat{\theta}^{(l)}$ of $\tilde{\theta}^{(l)}$, a plug-in estimator of the discrepancy
measure $d$ is
\begin{align*}
\hat{d}
=& \int_{\mathcal{X}} [ g_0(x^{\top} \hat{\theta}^{(0)}) -   
    g_1(x^{\top} \hat{\theta}^{(1)}) ]^2 d\hat{F}(x)\\
=&  \frac{1}{n}\sum_{i=1}^{n_0} \left[ g_0((\hat{\theta}^{(0)})^{\top} X_i^{(0)}) 
    -g_1( (\hat{\theta}^{(1)})^{\top} X_i^{(0)}) \right]^2 + 
    \frac{1}{n} \sum_{j=1}^{n_1} \left[ g_0(( \hat{\theta}^{(0)})^{\top} X_j^{(1)}) - g_1( (\hat{\theta}^{(1)})^{\top} X_j^{(1)}) \right]^2,
\end{align*}
where $n=n_0+n_1$ and $\hat{F}(x) = \frac{n_0}{n} \hat{F}_0(x)+ \frac{n_1}{n} \hat{F}_1(x) $ with $\hat{F}_l(x)$ being the empirical distribution functions of $\{ X^{(l)}_i \}_{i=1}^{n_l} $ for $l=0,1$.

We first investigate the asymptotic properties of $\hat{d}$ in high-dimensional settings, which provide the basis for the similarity testing. For this, we introduce some notions and regularity conditions. 
A random variable $W \in \mathbb{R}$ is called sub-Weibull of order $\tau > 0$, if
$ \|W\|_{\psi_{\tau}}:=\inf\{\eta>0: E[\psi_{\tau}(|W|/\eta)] \leq 1 \} < \infty $,
where $\psi_{\tau}(x) = \exp(x^{\tau}) - 1$ for $x \geq 0$.
According to the Markov inequality, if $W$ is sub-Weibull of order $\tau$, then
$ \mathbb{P}(|W| > t) \leq 2 \exp(-t^{\tau}/\|W\|_{\psi_{\tau}}^{\tau} )$
for all $t \geq 0$.
It is readily seen that sub-Gaussian and sub-exponential random variables are the special cases of sub-Weibull distributions with $\tau=2$ and $\tau=1$, respectively. More details on sub-Weibull distributions can be found in \cite{Vladimirova2020} and \cite{Kuchibhotla2022}. Let $\hat{S}^{(l)} = \{1 \leq j \leq p: \hat{\theta}_j^{(l)} \neq 0\}$ be the support of $\hat{\theta}^{(l)}$ and $\hat{s}^{(l)} = |\hat{S}^{(l)}|$.

\begin{condition}\label{A1}
{\rm (i)} The estimator $\hat{\theta}^{(l)}$ satisfies $\|\hat{\theta}^{(l)} - \tilde{\theta}^{(l)} \|_2 = O_p(\sqrt{\frac{\tilde{s}^{(l)} \log{p}}{n_l}})$ and $\| \hat{\theta}^{(l)} - \tilde{\theta}^{(l)} \|_1 = O_p( \tilde{s}^{(l)} \sqrt{\frac{\log{p}}{n_l}})$, where $l=0, 1$; {\rm (ii)} There exists a zero-mean random vector $L_l(X_i^{(l)}, Y_i^{(l)}, \tilde{\theta}^{(l)})$ such that 
\begin{align}\label{theta-expansion}
\sqrt{n_l}(\hat{\theta}^{(l)} - \tilde{\theta}^{(l)}) = \frac{1}{\sqrt{n_l}} \sum_{i=1}^{n_l} L_l(X_i^{(l)}, Y_i^{(l)}, \tilde{\theta}^{(l)}) + o_p(1).
\end{align}
\end{condition}

\begin{condition}\label{A2}
The function $g_l(\cdot)$ is twice differentiable and
$ |g_l(\theta^{\top}x)| \vee |\dot{g}_l(\theta^{\top}x)| \vee |\ddot{g}_l(\theta^{\top}x)| \leq B(x)$ 
for all $\theta \in \Theta_l$, where $l=0, 1$ and $B(X^{(l)})$ is sub-Weibull of order $\tau_0$ with $\| B(X^{(l)}) \|_{\psi_{\tau_0}} \leq C < \infty $. Moreover, the random variables $\epsilon^{(l)}$ and $\alpha^{\top}X^{(l)}$ are sub-Weibull of order $2$ with $ \| \epsilon^{(l)} \|_{\psi_2} \vee \| \alpha^{\top}X^{(l)} \|_{\psi_2} \leq C < \infty $ for all $ \alpha \in \mathcal{S}^{p-1} $.
\end{condition}

\begin{condition}\label{moment-condition}
$\mathbb{E} \{ |g_l( (\tilde{\theta}^{(l)})^{\top} X^{(l)})|^2 \} \leq C < \infty$ for $l=0,1$, $\mathbb{E} \{|g_1( (\tilde{\theta}^{(1)})^{\top} X^{(0)})|^2 \} \leq C < \infty$, and $\mathbb{E} \{|g_0( (\tilde{\theta}^{(0)})^{\top} X^{(1)})|^2 \} \leq C < \infty$.
\end{condition}

When the underlying regression models are correctly specified, Condition \ref{A1}(i) is satisfied by many penalized estimators, including the SCAD and MCP estimators, under suitable regularity conditions in high-dimensional settings. \citet{tan2026adf} further showed that similar rates hold for many penalized estimators under model misspecification. Lemma 6 of \cite{fan2004nonconcave} showed that Condition \ref{A1} (ii) holds for the SCAD estimator. Conditions \ref{A2} and \ref{moment-condition} are standard regularity assumptions in high-dimensional data analysis; see, for example, \cite{buhlmann2011} and \cite{tan2025b}.

The following theorem provides the asymptotic properties of $\hat{d}$ building on these conditions. We write $\psi(x,\theta^{(0)},\theta^{(1)})=g_0(x^{\top} \theta^{(0)}) - g_1(x^{\top} \theta^{(1)})$. 

\begin{theorem}\label{th:dhatasymnormal}
Suppose that Conditions \ref{A1}, \ref{A2}, and \ref{moment-condition} hold. If $\sqrt{n}(n_0/n - \zeta) = o(1)$, and $(\tilde{s}^{(0)} \vee \tilde{s}^{(1)})^2 (\log{p})^{2 \vee (4/3+2/\tau_0)} = o(n)$, then 
$$
\frac{\sqrt{n}(\hat{d}-d)}{\sigma_d} \stackrel{D}{\longrightarrow} \mathcal{N}(0,1),
$$
where the asymptotic variance $\sigma^2_{d} >0$ is
\begin{align*}
\sigma_{d}^2 
=& \zeta\, \mathbb{E} \Big\{ \psi^2(X^{(0)}, \tilde{\theta}^{(0)}, \tilde{\theta}^{(1)}) 
   -\mathbb{E}\{\psi^2(X^{(0)}, \tilde{\theta}^{(0)}, \tilde{\theta}^{(1)})\} \\
&   +2[M_{00} + (\zeta^{-1}-1) M_{10}]^{\top} L_0(X^{(0)}, Y^{(0)}, \tilde{\theta}^{(0)}) \Big\}^2 \\
&  + (1-\zeta)\, \mathbb{E} \Big\{ \psi^2(X^{(1)}, \tilde{\theta}^{(0)}, \tilde{\theta}^{(1)}) 
   - \mathbb{E}\{\psi^2(X^{(1)}, \tilde{\theta}^{(0)}, \tilde{\theta}^{(1)})\} \\
&- 2\Big[ M_{01}/(\zeta^{-1}-1) + M_{11} \Big]^{\top} L_1(X^{(1)}, Y^{(1)}, \tilde{\theta}^{(1)}) \Big\}^2,
\end{align*}
with
\begin{align*}
& M_{00} = \mathbb{E}\big[ \psi(X^{(0)}, \tilde{\theta}^{(0)}, \tilde{\theta}^{(1)}) X^{(0)} 
           \dot{g}_0((X^{(0)})^{\top} \tilde{\theta}^{(0)}) \big] \\
& M_{01} = \mathbb{E}\big[ \psi(X^{(0)}, \tilde{\theta}^{(0)}, \tilde{\theta}^{(1)}) X^{(0)} 
           \dot{g}_1((X^{(0)})^{\top} \tilde{\theta}^{(1)}) \big] \\
& M_{10} = \mathbb{E}\big[ \psi(X^{(1)}, \tilde{\theta}^{(0)}, \tilde{\theta}^{(1)}) X^{(1)} 
           \dot{g}_0((X^{(1)})^{\top} \tilde{\theta}^{(0)}) \big] \\
& M_{11} = \mathbb{E}\big[ \psi(X^{(1)}, \tilde{\theta}^{(0)}, \tilde{\theta}^{(1)}) X^{(1)} 
           \dot{g}_1((X^{(1)})^{\top} \tilde{\theta}^{(1)}) \big].
\end{align*}
\end{theorem}

Since the asymptotic variance lies in the denominator in Theorem~\ref{th:dhatasymnormal}, we require $\sigma_d^2 > 0$ and exclude the degenerate case $\sigma^2_d=0$. Let $\hat{\sigma}^2_d$ be a consistent estimator of $\sigma^2_d$. We then obtain the limiting distribution of $\sqrt{n}(\hat{d}-d)/\hat{\sigma}_d$.
\begin{corollary}
Suppose that the conditions of Theorem~\ref{th:dhatasymnormal} hold. Then
$
\sqrt{n}(\hat{d}-d)/\hat{\sigma}_d \stackrel{D}{\longrightarrow} \mathcal{N}(0,1).
$
\end{corollary}

\subsection{Motivation from the LFC-based approach} 
For testing $H_0^{(\Delta)}: 0 < d \leq \Delta$, a traditional and widely adopted construction in the literature is to define the test statistic under the least favorable configuration (LFC). This strategy relies exclusively on the boundary value $d = \Delta$ to define the statistic, ensuring asymptotic type I error control for every fixed null parameter value $d \leq \Delta$. Specifically, the LFC-based test statistic for $H_0^{(\Delta)}$ is defined as
\begin{align}\label{test:LFC}
\hat T^{\rm LFC} = \frac{\sqrt{n}(\hat{d} - \Delta)}{\hat{\sigma}_{d}}.
\end{align}
By Theorem \ref{th:dhatasymnormal}, when $d = \Delta$, $\hat{T}^{\rm LFC}$ is asymptotically standard normal. Accordingly, we reject \(H_0^{(\Delta)}\) whenever $\hat{T}^{\rm LFC} \ge z_{1-\alpha}$, where $z_{1-\alpha}$ denotes the $1-\alpha$-quantile of the standard normal distribution. 
However, the LFC approach suffers from the limitations of over-conservatism. Since the LFC method accounts for the so-called ``worst case'' to ensure Type I error control under $H_0^{(\Delta)}$, it rigidly anchors the test to the boundary $\Delta$, disregarding the actual position of the true discrepancy $d$ under the null $H_0^{(\Delta)}$. The asymptotic rejection probability is
\begin{align*}
\lim_{n\to \infty}\mathbb{P}\left(\frac{\sqrt{n}(\hat{d}-\Delta)}{\hat{\sigma}_{d}} \ge z_{1-\alpha} \right) = \lim_{n\to \infty}\mathbb{P}\left(\frac{\sqrt{n}(\hat{d}-d)}{\hat{\sigma}_{d}} \ge z_{1-\alpha} + \frac{\sqrt{n}(\Delta-d)}{\hat{\sigma}_{d}} \right).
\end{align*} 
If $d < \Delta$, the discrepancy term ${\sqrt{n}(\Delta - d)}/{\hat{\sigma}_d}$  diverges to infinity, causing the Type I error to approach zero. The supremum of the limiting probabilities over all $d\le \Delta$ equals the significance level $\alpha$. In other words, while this strategy ensures the test is ``safe'', it may be overly conservative at any fixed interior point of the null $ H_0^{(\Delta)} $. 

From the perspective of power, the LFC-based test $\hat{T}^{\rm LFC}$ has a natural locally asymptotically uniformly most powerful (LAUMP) interpretation.
Consider local alternatives of the form \( d=\Delta+\frac{h}{\sqrt n} \) with \( h >0 \).
The asymptotic normality of $\hat d$ implies that 
\[
\hat T^{\rm LFC}
= \frac{\sqrt n(\hat d-\Delta)}{\hat\sigma_d}
\stackrel{D}{\longrightarrow}
N\left(\frac{h}{\sigma_d},1\right).
\]
Thus, the limiting testing problem for similarity becomes the Gaussian shift experiment
\[
H_0^{S}:h\leq 0
\qquad \text{versus} \qquad
H_1^{S}:h>0.
\]
In this limiting experiment, the rejection rule of the LFC-based test is \( Z\geq z_{1-\alpha} \), where $Z \sim N(h/\sigma_d, 1)$. 
For this Gaussian location family, the likelihood ratio of \(N(h/\sigma_d,1)\)
relative to \(N(0,1)\) is 
\[ \frac{\phi(z-h/\sigma_d)}{\phi(z)} = \exp \left( \frac{h}{\sigma_d} z -  \frac{h^2}{2\sigma_d^2} \right), \quad h > 0, \]
where $\phi(\cdot)$ denotes the density function of the standard normal distribution. 
Since this likelihood ratio is monotone increasing in $z$, it follows from the Neyman-Pearson lemma or the Karlin--Rubin theorem \citep{karlin1956theory} that the rejection rule \( Z\geq z_{1-\alpha} \) is uniformly most powerful (UMP) in the limiting testing problem.
Consequently, the LFC-based test $\hat T^{\rm LFC}$ attains the local asymptotic power envelope
\[
1-\Phi\left(z_{1-\alpha}-\frac{h}{\sigma_d}\right),
\qquad h>0,
\]
where $\Phi(\cdot)$ denotes the cumulative distribution function of the standard normal distribution. 
Thus, although the LFC-based test $\hat T^{\rm LFC}$ is conservative for fixed interior null points $d<\Delta$, its local asymptotic power near the boundary $d=\Delta$ cannot be improved by any level-$\alpha$ test whose statistic has the same limiting distribution $N(h/\sigma_d,1)$. We refer to this as the locally asymptotically uniformly most powerful (LAUMP) property of the LFC-based test.  
We therefore propose an adaptive boundary-proxy test that avoids the over-conservatism of LFC-based methods and simultaneously retains the LAUMP property.

\subsection{Adaptive boundary-proxy test for similarity}\label{sec:ada} 
In this subsection, we propose an adaptive boundary-proxy test designed to attain the nominal significance level at every fixed configuration of $H_0^{(\Delta)}$ while retaining the LAUMP property of the LFC-based test. Our approach is motivated by the decomposition: 
\begin{align}\label{dec}
\hat{d} - \Delta = \underbrace{\hat{d} - d}_{\text{Estimation Error}} + \underbrace{d - \Delta}_{ \text{Boundary Offset}},
\end{align} 
where $d - \Delta$ acts as a positive shift under the alternative $H_1^{(\Delta)}$. Because $d$ is unknown, we employ a sample-splitting technique, partitioning the data into two independent subsamples, $\mathcal{D}_1$ and $\mathcal{D}_2$, of equal size to compute estimators $\hat{d}_1$ and $\hat{d}_2$ of the discrepancy $d$. It follows from Theorem \ref{th:dhatasymnormal} that
$$ \sqrt{n/2}(\hat{d}_k-d)/\hat{\sigma}_d \stackrel{D}{\longrightarrow} \mathcal{N}(0,1),
\quad {\rm for} \ k=1, 2.
$$
According to the decomposition~(\ref{dec}), we find that the estimation error $\hat{d}-d$ is asymptotically normal under both the null and the alternative and the shift term $d-\Delta$ is used to achieve high power under the alternatives. Thus, our purpose is to construct an adaptive test that retains only the estimation error term $\hat{d}-d$ under the null and retains only the shift term $d - \Delta$ under the alternative. To achieve this, we first define the Error Proxy ($\hat{T}_c$) that captures the centered estimation error $\hat{d} - d$: 
$$ \hat{T}_{c} = \frac{ \sqrt{n} (\hat{d}_1 - \hat{d}_2) }{ 2\hat{\sigma}_{d} }
= \frac{\sqrt{n/2}(\hat{d}_1-d)}{\sqrt{2}\hat{\sigma}_d} - \frac{\sqrt{n/2}(\hat{d}_2-d)}{\sqrt{2}\hat{\sigma}_d}. $$ 
By Theorem \ref{th:dhatasymnormal} and the independence between $\hat{d}_1$ and $\hat{d}_2$, $\hat{T}_c$ converges to $\mathcal{N}(0,1)$ in distribution under both the null and alternative, and this proxy provides a stable distribution for the critical value.
To inherit the LAUMP property of the LFC-based test, we employ the test statistic $\hat{T}^{\mathrm{LFC}}$ defined in~(\ref{test:LFC}) as the second part of the adaptive boundary-proxy test that captures the shift term $d - \Delta$.
Therefore, the adaptive boundary-proxy test is defined as
\begin{align}\label{adaptive-test}
\hat{T}^{\text{ada}} = \hat{T}_c \mathbb{I}(\hat{T}^{\rm LFC} \leq -a_n) + \hat{T}^{\rm LFC} \mathbb{I}(\hat{T}^{\rm LFC} > -a_n),
\end{align}
where $a_n$ is a positive sequence satisfying $a_n \to \infty$ and $a_n/\sqrt{n} \to 0$. 

Under $H_0^{(\Delta)}$, $\hat{T}^{\rm LFC}$ diverges to negative infinity at the rate $\sqrt{n}$ if $d < \Delta$, and converges in distribution to $\mathcal{N}(0,1)$ if $d = \Delta$. Under $H_1^{(\Delta)}$ ($d > \Delta$), it diverges to infinity at the rate $\sqrt{n}$. Since $a_n/\sqrt{n} \to 0$, under $H_0^{(\Delta)}$ with $d < \Delta$, we have $ \hat{T}_c \mathbb{I}(\hat{T}^{\mathrm{LFC}} > -a_n) = o_p(1) $ and $ \hat{T}^{\mathrm{LFC}} \mathbb{I}(\hat{T}^{\mathrm{LFC}} > -a_n) = o_p(1)$, which yields
$$ \hat{T}^{\mathrm{ada}} = \hat{T}_c + o_p(1) \stackrel{D}{\longrightarrow} \mathcal{N}(0,1). $$
Under $H_0^{(\Delta)}$ with $d = \Delta$ or under $H_1^{(\Delta)}$ with $d > \Delta$, since $a_n \to \infty$, we have
$ \hat{T}_c \mathbb{I}(\hat{T}^{\mathrm{LFC}} \leq -a_n) = o_p(1) $
and $ \hat{T}^{\mathrm{LFC}} \mathbb{I}(\hat{T}^{\mathrm{LFC}} \leq -a_n) = o_p(1)$.
Consequently, under $H_0^{(\Delta)}$ with $d = \Delta$,
$$ \hat{T}^{\mathrm{ada}} = \hat{T}^{\mathrm{LFC}} + o_p(1) \stackrel{D}{\longrightarrow} \mathcal{N}(0,1).$$
Under $H_1^{(\Delta)}$ with $d > \Delta$, $ \hat{T}^{\mathrm{ada}} = \hat{T}^{\mathrm{LFC}} + o_p(1)$, which diverges to infinity at rate $\sqrt{n}$.

We further consider the local alternatives for the similarity-testing problem:
\begin{align*}
    H_{1n}^{(\Delta)}: d = \Delta + \delta_{n_2},
\end{align*}
where $\delta_{n_2}$ is a positive sequence tending to zero. 
The following theorem summarizes the asymptotic properties of $\hat{T}^{\rm ada}$ under the null and various alternative hypotheses. 

\begin{theorem}\label{th2}
Assume that $\sigma_d^2>0$, $a_n \to \infty$, $a_n/\sqrt{n} \to 0$, and that Conditions \ref{A1}, \ref{A2}, and \ref{moment-condition} hold. 

{\rm (i)} Under the null $H_0^{(\Delta)}$, $ \hat{T}^{\rm ada} \stackrel{D}{\longrightarrow} \mathcal{N}(0, 1). $

{\rm (ii)} Under the local alternative $H_{1n}^{(\Delta)} $ with $\delta_{n_2} = C/\sqrt{n}$ for some constant $ C > 0$,
$ \hat{T}^{\rm ada} \stackrel{D}{\longrightarrow} \mathcal{N}(C/\sigma_d, 1). $

{\rm (iii)} Under the local alternative $H_{1n}^{(\Delta)} $ with $\sqrt{n} \delta_{n_2} \to \infty$,
$ \frac{1}{\sqrt{n} \delta_{n_2}} \hat{T}^{\rm ada} \stackrel{P}{\longrightarrow} \frac{1}{\sigma_d}. $

{\rm (iv)} Under the global alternative $H_1^{(\Delta)}$,
$ \frac{1}{\sqrt{n}} \hat{T}^{\rm ada} \stackrel{P}{\longrightarrow} \frac{d - \Delta}{\sigma_d}.$
\end{theorem}

By Theorem~\ref{th2}, under each fixed null hypothesis $H_0^{(\Delta)}$, we have $\mathbb{P}(\hat{T}^{\rm ada} > z_{1-\alpha}) \to \alpha$. This implies that the proposed adaptive test $\hat{T}^{\rm ada}$ can maintain the nominal significance level at each fixed point in $H_0^{(\Delta)}$, thereby addressing the conservatism of the traditional LFC-based test $\hat{T}^{\mathrm{LFC}}$. 
Under $H_1^{(\Delta)}$, the proposed adaptive test is asymptotically equivalent to the LFC-based test $\hat{T}^{\mathrm{LFC}}$. Thus, our proposed test inherits the LAUMP property of the LFC-based test. Furthermore, we can show that, under the alternative $H_1^{(\Delta)}$, the proposed adaptive test $\hat{T}^{\mathrm{ada}}$ is at least as powerful as the LFC-based test $\hat{T}^{\mathrm{LFC}}$ in finite samples. Indeed, for a nominal significance level $0 < \alpha < 0.5$,
\begin{align*}
\mathbb{P}(\hat{T}^{\mathrm{ada}} > z_{1-\alpha}) 
&=  \mathbb{P}(\hat{T}^{\mathrm{ada}} > z_{1-\alpha}, \hat{T}^{\mathrm{LFC}} \leq -a_n) +
    \mathbb{P}(\hat{T}^{\mathrm{ada}} > z_{1-\alpha}, \hat{T}^{\mathrm{LFC}} > -a_n) \\
&=  \mathbb{P}(\hat{T}_c > z_{1-\alpha}, \hat{T}^{\mathrm{LFC}} \leq -a_n) +
    \mathbb{P}(\hat{T}^{\mathrm{LFC}} > z_{1-\alpha}, \hat{T}^{\mathrm{LFC}} > -a_n) \\
&=  \mathbb{P}(\hat{T}_c > z_{1-\alpha}, \hat{T}^{\mathrm{LFC}} \leq -a_n) +
    \mathbb{P}(\hat{T}^{\mathrm{LFC}} > z_{1-\alpha}) \\
&\geq \mathbb{P}(\hat{T}^{\mathrm{LFC}} > z_{1-\alpha}),
\end{align*}
where the third equation follows from $z_{1-\alpha} >0$.
This implies that our proposed test is at least as powerful as the LFC-based test $\hat{T}^{\rm LFC}$ for every finite sample size.

\begin{remark}
Another approach in the literature seeking to mitigate the limitation of the least favorable configuration in inequality-constrained testing is the recentering method, which was first proposed by \cite{hansen2005test} in the context of testing for superior predictive ability; see also \citet{linton2010improved},  \citet{AndrewsSoares2010}, \citet{donald2011new}, and \citet{donald2016improving}. However, by using a truncated critical value, such methods can maintain the significance level only on the boundary of the composite null hypothesis; see, for example, Theorem 2 of \citet{linton2010improved} or Theorem 4.1 of \citet{donald2016improving}. In contrast, the proposed adaptive boundary-proxy test attains the nominal level at every fixed null configuration.
\end{remark}

\section{Adaptive-to-Model Test for Similarity and Model Misspecification}\label{sec:adam}
\subsection{Test construction and null distribution}
We construct an adaptive-to-model test to simultaneously assess functional similarity and model specification for the two populations. Recall that the joint null hypothesis $H_0$ is equivalent to the intersection $H_0^{(0)} \cap H_0^{(1)} \cap H_0^{(\Delta)}$, where 
\begin{align*}
& H_0^{(0)}: \mathbb{P}\{ \mathbb{E}[Y^{(0)}|X^{(0)}] = g_0((\theta_0^{(0)})^{\top}X^{(0)} ) \} = 1, 
 \quad  {\rm for \ some} \ \theta_0^{(0)} \in \Theta_0,  \\
& H_0^{(1)}: \mathbb{P}\{ \mathbb{E}[Y^{(1)}|X^{(1)}] = g_1( (\theta_0^{(1)})^{\top}X^{(1)} )\} = 1, 
 \quad  {\rm for \ some} \ \theta_0^{(1)} \in \Theta_1,  \\
& H_0^{(\Delta)}: 0 < d \leq \Delta.
\end{align*}
The principle of the adaptive-to-model testing framework for assessing $H_0$ is similar to that of the adaptive test $\hat{T}^{\rm ada}$ for testing the similarity of two populations. 
Specifically, we first construct test statistics for three component null hypotheses $H_{0}^{(0)}$, $H_{0}^{(1)} $, and $H_{0}^{(\Delta)}$. Let $\hat{T}_l^{\rm mod}$ be the test statistic for assessing the component null hypothesis $H_0^{(l)}$, where $l=0, 1$. 
Suppose that the test statistic $\hat{T}_l^{\rm mod}$ converges in distribution to a fixed, nondegenerate limiting distribution under $H_0^{(l)}$ and diverges to infinity in probability at rate $r_{n_l}$ under $H_1^{(l)}$. 
We then propose the following adaptive-to-model test for $H_0$:
\begin{align}\label{adam-test}
\begin{split}
\hat{T}^{\rm adam}  
=&  \hat{T}^{\rm ada} + |\hat{T}_0^{\rm mod}|\mathbb{I}(|\hat{T}_0^{\rm mod}| 
    > b_{n_0}) + |\hat{T}_1^{\rm mod}|\mathbb{I}(|\hat{T}_1^{\rm mod}| > b_{n_1}),  
\end{split}
\end{align}
where $\hat{T}^{\rm ada}$ is given in (\ref{adaptive-test}) and $b_{n_l}$ is a positive sequence satisfying that $b_{n_l} \to \infty$ and $b_{n_l}/r_{n_l} \to 0$, where $l=0, 1$.

As with the tuning parameter $a_n$ used in the construction of $\hat T^{\rm ada}$, the tuning parameters $b_{n_l}$ are introduced to ensure that the test $\hat{T}^{\rm adam}$ satisfies the following adaptive-to-model properties. Note that the joint null hypothesis $H_0$ is equivalent to $H_0^{(0)}$, $H_0^{(1)}$ and $H_0^{(\Delta)}$ holding simultaneously.
Since $b_{n_l} \to \infty$ and $\hat{T}_l^{\rm mod}$ converges in distribution to a nondegenerate distribution under $H_0^{(l)}$, 
$ |\hat{T}_l^{\mathrm{mod}}| \mathbb{I}(|\hat{T}_l^{\rm mod}| > b_{n_l}) = o_p(1)
$
for $l=0, 1$. Consequently, under $H_0$, we obtain
$ \hat{T}^{\rm adam} = \hat{T}^{\rm ada} + o_p(1). $
This implies that the test $\hat{T}^{\rm adam}$ is asymptotically equivalent to $\hat{T}^{\rm ada}$ under $H_0$. Let $0 < \alpha < 0.5$ be the nominal significance level. 

\begin{theorem}\label{T_adam_H0}
Assume that Conditions~\ref{A1}, \ref{A2}, and \ref{moment-condition} hold. If $\sigma_d^2>0$, $a_n \to \infty$, $a_n/\sqrt{n} \to 0$, and $b_{n_l} \to \infty$ for $l=0,1$, then, under $H_0$,
$ \hat{T}^{\rm adam} \stackrel{D}{\longrightarrow} \mathcal{N}(0,1). $
Consequently, $\lim_{n \to \infty} \mathbb{P}(\hat{T}^{\rm adam} > z_{1-\alpha}) = \alpha$.
\end{theorem}

Theorem~\ref{T_adam_H0} implies that the adaptive-to-model test $\hat{T}^{\rm adam}$ can asymptotically maintain the significance level at every fixed null configuration in $H_0$, and thus avoids the over-conservatism of the LFC-based tests. 

For practical use, we need to specify the tests for the component null hypotheses $H_0^{(0)}$ and $H_0^{(1)}$. There are limited studies in the literature for model checking that are applicable in
high-dimensional settings in which the predictor dimension $p$ may far exceed the sample size $n$. 
In this paper, we adopt the Fisher-combination-based test proposed by \cite{tan2025b} which assesses the goodness-of-fit of sparse parametric regression models in high-dimensional settings. Their test statistic is defined as
\begin{align}\label{fisher}
\hat{T}_{\text{Fisher}}^{(l)} = -2 \sum_{k=1}^{d_0} \ln(\hat{p}_{\alpha_{lk}}^{(l)}),
\end{align}
where $l=0, 1$ and $\hat{p}_{\alpha_{lk}}^{(l)} = 1-\Phi(\hat{T}_{\alpha_{lk}}^{(l)})$. Here, the projections $\alpha_{l1}, \dots, \alpha_{l d_0}$ belong to the unit sphere $\mathcal{S}^{p-1}$, and 
$$
\hat{T}_{\alpha_{lk}}^{(l)}
= \frac{\sum_{1\leq i \neq j \leq n_l} \varepsilon_i^{(l)}(\hat{\theta}^{(l)})
  \varepsilon^{(l)}_j(\hat{\theta}^{(l)})K(\frac{\alpha_{lk}^{\top} X_i^{(l)} - \alpha_{lk}^{\top}X_j^{(l)}}{h})}{\left(2 \sum_{1\leq i \neq j \leq n_l} \Big[ \varepsilon_i^{(l)}(\hat{\theta}^{(l)}) \varepsilon_j^{(l)}(\hat{\theta}^{(l)}) K(\frac{\alpha_{lk}^{\top}X_i^{(l)} - \alpha_{lk}^{\top}X_j^{(l)}}{h}) \Big]^2 \right)^{1/2}},
$$
where $\varepsilon_i^{(l)}(\hat{\theta}^{(l)}) = Y_i^{(l)} - g_l((\hat{\theta}^{(l)})^{\top}X_i^{(l)})$ denotes the residual for observation $i$ in population $l$, $K(\cdot)$ is the kernel function, and $h$ is the bandwidth. For the choice of the projections $\alpha_{l1}, \dots, \alpha_{ld_0}$, we refer the readers to Section 4 of \citet{tan2025b}.

\subsection{Power analysis of the adaptive-to-model test $\hat{T}^{\rm adam}$}
In this subsection, we investigate the asymptotic properties of $\hat{T}^{\rm adam}$ under various alternative hypotheses. 
Recall that the alternative hypothesis $H_1$ is equivalent to the case where at least one of the component null hypotheses $H_{0}^{(0)}$, $H_{0}^{(1)} $, and $H_{0}^{(\Delta)}$ fails to hold. Elementary calculations show that $H_1$ can be partitioned into $7$ mutually exclusive sub-alternative hypotheses: 
\begin{align}\label{alter}
\begin{split}
& H_{11} = H_1^{(0)} \cap H_0^{(1)} \cap H_0^{(\Delta)}, \quad
  H_{12} = H_0^{(0)} \cap H_1^{(1)} \cap H_0^{(\Delta)}, \\
& H_{13} = H_1^{(0)} \cap H_1^{(1)} \cap H_0^{(\Delta)}, \quad
  H_{14} = H_0^{(0)} \cap H_0^{(1)} \cap H_1^{(\Delta)}, \\
& H_{15} = H_1^{(0)} \cap H_0^{(1)} \cap H_1^{(\Delta)}, \quad
  H_{16} = H_0^{(0)} \cap H_1^{(1)} \cap H_1^{(\Delta)}, \\
& H_{17} = H_1^{(0)} \cap H_1^{(1)} \cap H_1^{(\Delta)}.
\end{split}
\end{align}
Next, we formulate the local alternative hypothesis $H_{1n}$ that converges to the joint null hypothesis $H_0$. To this end, define
\begin{align*}
& H_{1n}^{(0)}: \mathbb{E}[Y^{(0)}|X^{(0)}]= g_0((\theta_0^{(0)})^{\top}X^{(0)})
              + \delta_{n_0} w_0(X^{(0)}), \quad a.s., \\
& H_{1n}^{(1)}: \mathbb{E}[Y^{(1)}|X^{(1)}]=g_1( ({\theta}_0^{(1)})^{\top}X^{(1)} ) 
              + \delta_{n_1} w_1(X^{(1)}), \quad a.s., \\
& H_{1n}^{(\Delta)}: d = \Delta + \delta_{n_2},
\end{align*}
where $\{ \delta_{n_i}\}_{i=0,1,2}$ are positive sequences tending to zero.
Analogously to the global alternative $H_1$, the local alternative hypothesis $H_{1n}$ can also be partitioned into the following $7$ mutually exclusive local sub-alternatives:
\begin{align}\label{local}
\begin{split}
& H_{11n} = H_{1n}^{(0)} \cap H_0^{(1)} \cap H_0^{(\Delta)}, \quad
  H_{12n} = H_0^{(0)} \cap H_{1n}^{(1)} \cap H_0^{(\Delta)}, \\
& H_{13n} = H_{1n}^{(0)} \cap H_{1n}^{(1)} \cap H_0^{(\Delta)}, \quad
  H_{14n} = H_0^{(0)} \cap H_0^{(1)} \cap H_{1n}^{(\Delta)}, \\
& H_{15n} = H_{1n}^{(0)} \cap H_0^{(1)} \cap H_{1n}^{(\Delta)}, \quad
  H_{16n} = H_0^{(0)} \cap H_{1n}^{(1)} \cap H_{1n}^{(\Delta)}, \\
& H_{17n} = H_{1n}^{(0)} \cap H_{1n}^{(1)} \cap H_{1n}^{(\Delta)}.
\end{split}
\end{align}

To derive the asymptotic properties of the adaptive-to-model test $\hat{T}^{\rm adam}$ under the various alternative hypotheses, we further impose the following regularity condition. 

\begin{condition}\label{assum-Tmod}
{\rm (i)} Under $H_1^{(l)}$,  $ r_{n_l}^{-1} \hat{T}_l^{\rm mod} \stackrel{P}{\longrightarrow} \mu_{1}^{(l)} $ for $l=0, 1$, where $\mu_{1}^{(l)} $ is a nonzero constant.
{\rm (ii)} Under $H_{1n}^{(l)}$ with $\delta_{n_l} \to 0$ and $r_{n_l} \delta_{n_l}^4 \to \infty $, 
$r_{n_l}^{-1} \delta^{-4}_{n_l} \hat{T}^{\rm mod}_l \stackrel{P}{\longrightarrow} \mu_{\Delta}^{(l)}$ 
for $l=0, 1$, where $\mu_{\Delta}^{(l)}$ is a nonzero constant.
{\rm (iii)} Under $H_{1n}^{(l)}$ with $\delta_{n_l} = r_{n_l}^{-1/4}$ as $n \rightarrow \infty$,
$ \hat{T}^{\rm mod}_l  \stackrel{D}{\longrightarrow} T_l $
for $l=0, 1$, where $T_l$ follows a fixed, nondegenerate distribution which differs from the limiting distribution of $\hat{T}^{\rm mod}_l$ under $H_0^{(l)}$. 
\end{condition}

It follows from \citet{tan2025b} that the Fisher-combination-based test $\hat{T}^{(l)}_{\rm Fisher}$ given in \eqref{fisher} satisfies Condition~\ref{assum-Tmod} with $r_{n_l} = n_l^2 h$, where $h \to 0$ is the bandwidth used in the test statistic $\hat{T}_{\alpha_{lk}}^{(l)}$. The following proposition summarizes the adaptive property of the model checking component for the test $\hat{T}^{\rm adam}$.

\begin{proposition}\label{prop:model-gating}
Suppose that $b_{n_l}\to\infty$, $b_{n_l}/r_{n_l}=o(1)$, and Condition~\ref{assum-Tmod} holds. Then the following results hold.

{\rm (1)} Under $H_1^{(l)}$, $|\hat T_l^{\rm mod}| \mathbb{I}(|\hat T_l^{\rm mod}| \leq b_{n_l}) = o_p(1)$ and thus,
$ |\hat T_l^{\rm mod}| \mathbb{I}(|\hat T_l^{\rm mod}| > b_{n_l}) = |\hat T_l^{\rm mod}| + o_p(1). $

{\rm (2)} Under $H_{1n}^{(l)}$ with $r_{n_l} \delta^4_{n_l} |\mu_{\Delta}^{(l)}|/b_{n_l} \ge C > 1$ for $n$ large enough, $ |\hat T_l^{\rm mod}| \mathbb{I}(|\hat T_l^{\rm mod}| \leq b_{n_l}) = o_p(1) $ and thus,
$ |\hat T_l^{\rm mod}| \mathbb{I}(|\hat T_l^{\rm mod}| > b_{n_l}) = |\hat T_l^{\rm mod}| + o_p(1). $

{\rm (3)} Under $H_{1n}^{(l)}$ with $r_{n_l} \delta^4_{n_l}|\mu_{\Delta}^{(l)}|/b_{n_l} \le C < 1$ for $n$ large enough, 
$ |\hat T_l^{\rm mod}| \mathbb{I}(|\hat T_l^{\rm mod}| > b_{n_l}) = o_p(1). $
\end{proposition}

Under $H_{1n}^{(l)}$ with $\delta_{n_l} = r_{n_l}^{-1/4}$, since $b_{n_l} \to \infty$, it follows that $ r_{n_l} \delta^4_{n_l}|\mu_{\Delta}^{(l)}|/b_{n_l} \le C < 1$ for $n_l$ large enough. Consequently, under $H_{1n}^{(l)}$ with $\delta_{n_l} = r_{n_l}^{-1/4}$, 
$ |\hat T_l^{\rm mod}| \mathbb{I}(|\hat T_l^{\rm mod}| > b_{n_l}) = o_p(1). $

Under the alternative $H_1$, at least one of the component null hypotheses $H_0^{(0)}$, $H_0^{(1)}$ and $H_0^{(\Delta)}$ is false. We then show that our test statistic $\hat{T}^{\rm adam}$ can adaptively retain the test statistic corresponding to each component null hypothesis that fails to hold.
For example, suppose that $H_{11}$ holds. Then $H_0^{(0)}$ is false, whereas both $H_0^{(1)}$ and $H_0^{(\Delta)}$ hold. Consequently, $ \hat{T}^{\rm ada} \stackrel{D}{\longrightarrow}  \mathcal{N}(0,1)$, $ |\hat{T}_0^{\rm mod}|\mathbb{I}(|\hat{T}_0^{\rm mod}| > b_{n_0}) = |\hat{T}_0^{\rm mod}| + o_p(1)$, and $ |\hat{T}_1^{\rm mod}|\mathbb{I}(|\hat{T}_1^{\rm mod}| > b_{n_1}) = o_p(1) $.
This yields $ \hat{T}^{\rm adam} = |\hat{T}_0^{\rm mod}| + O_p(1)$. 
Thus, $\hat{T}^{\rm adam}$ adaptively retains $\hat{T}_0^{\rm mod}$ when the corresponding component null hypothesis $H_0^{(0)}$ fails.
Since $\hat{T}_0^{\rm mod}$ diverges to infinity in probability at the rate $r_{n_0}$ under $H_1^{(0)}$, $\hat{T}^{\rm adam}$ also diverges to infinity in probability at the same rate.
Consequently, the proposed test $\hat{T}^{\rm adam}$ is consistent and has asymptotic power one. 
Analogous arguments apply to the other configurations under $H_1$. 
The more specific asymptotic properties of \(\hat{T}^{\rm adam}\) under various alternative hypotheses are summarized in the following theorem.

\begin{theorem}\label{power-adam}
Suppose that $\sigma_d^2>0$, $a_n \to \infty$, $a_n/\sqrt{n} \to 0$, $b_{n_l} \to \infty$ and $b_{n_l}/r_{n_l} \to 0$ for $l=0,1$, and that Conditions \ref{A1}, \ref{A2}, \ref{moment-condition} and \ref{assum-Tmod} hold.  

{\rm (1)} Under $H_{11}$, $H_{12}$, $H_{13}$, $H_{14}$, $H_{15}$, $H_{16}$, and $H_{17}$, $\hat T^{\rm adam}$ diverges to infinity in probability  at the rates
$r_{n_0}, r_{n_1}, r_{n_0}\vee r_{n_1}, \sqrt{n}, r_{n_0}\vee\sqrt{n}, r_{n_1}\vee\sqrt{n}, r_{n_0} \vee r_{n_1} \vee \sqrt{n}$, 
respectively.

{\rm (2)} Under $H_{11n}$ with $r_{n_0} \delta^4_{n_0} |\mu_{\Delta}^{(0)}|/b_{n_0} \ge C > 1$ for $n$ large enough,  $\hat{T}^{\rm adam}$ diverges to infinity in probability at rate $r_{n_0} \delta_{n_0}^4$;
under $H_{11n}$ with $r_{n_0} \delta^4_{n_0} |\mu_{\Delta}^{(0)}|/b_{n_0} \le C < 1$ for $n$ large enough, 
$ \hat{T}^{\rm adam} \stackrel{D}{\longrightarrow} \mathcal{N}(0,1). $

{\rm (3)} Under $H_{12n}$ with $r_{n_1} \delta^4_{n_1}|\mu_{\Delta}^{(1)}|/b_{n_1} \ge C > 1$ for $n$ large enough, 
$\hat{T}^{\rm adam}$ diverges to infinity in probability at rate $r_{n_1} \delta_{n_1}^4$;
under $H_{12n}$ with $r_{n_1} \delta^4_{n_1} |\mu_{\Delta}^{(1)}|/b_{n_1} \le C < 1$ for $n$ large enough, 
$ \hat{T}^{\rm adam} \stackrel{D}{\longrightarrow} \mathcal{N}(0,1). $

{\rm (4)} Under $H_{13n}$ with $r_{n_0} \delta^4_{n_0} |\mu_{\Delta}^{(0)}|/b_{n_0} \ge C > 1$ or $r_{n_1} \delta^4_{n_1} |\mu_{\Delta}^{(1)}|/b_{n_1} \ge C > 1$ for $n$ large enough, 
$\hat{T}^{\rm adam}$ diverges to infinity in probability at least at rate $r_{n_0} \delta_{n_0}^4 \wedge r_{n_1} \delta_{n_1}^4$;
under $H_{13n}$ with $r_{n_0} \delta^4_{n_0}|\mu_{\Delta}^{(0)}|/b_{n_0} \le C < 1$ and $r_{n_1} \delta^4_{n_1} |\mu_{\Delta}^{(1)}|/b_{n_1} \le C < 1$ for $n$ large enough, 
$ \hat{T}^{\rm adam} \stackrel{D}{\longrightarrow} \mathcal{N}(0,1). $

{\rm (5)} Under $H_{14n}$ with $\sqrt{n} \delta_{n_2} \to \infty$, 
$\hat{T}^{\rm adam}$ diverges to infinity in probability at rate $\sqrt{n} \delta_{n_2}$;
under $H_{14n}$ with $\delta_{n_2} = C/\sqrt{n}$ for some constant $C > 0$, 
$ \hat{T}^{\rm adam} \stackrel{D}{\longrightarrow} \mathcal{N}(C/\sigma_d, 1). $

{\rm (6)} Under $H_{15n}$ with $r_{n_0} \delta^4_{n_0} |\mu_{\Delta}^{(0)}|/b_{n_0} \ge C > 1$ for $n$ large enough or $\sqrt{n} \delta_{n_2} \to \infty$, 
$\hat{T}^{\rm adam}$ diverges to infinity in probability at least at rate $r_{n_0} \delta_{n_0}^4 \wedge \sqrt{n} \delta_{n_2}$;
under $H_{15n}$ with $r_{n_0} \delta^4_{n_0} |\mu_{\Delta}^{(0)}|/b_{n_0} \le C < 1$ for $n$ large enough and  $\delta_{n_2} = C/\sqrt{n}$, 
$ \hat{T}^{\rm adam} \stackrel{D}{\longrightarrow} \mathcal{N}(C/\sigma_d, 1). $

{\rm (7)} Under $H_{16n}$ with $r_{n_1} \delta^4_{n_1} |\mu_{\Delta}^{(1)}|/b_{n_1} \ge C > 1$ for $n$ large enough or $\sqrt{n} \delta_{n_2} \to \infty$, 
$\hat{T}^{\rm adam}$ diverges to infinity in probability at least at rate $r_{n_1} \delta_{n_1}^4 \wedge \sqrt{n} \delta_{n_2}$;
under $H_{16n}$ with $r_{n_1} \delta^4_{n_1} |\mu_{\Delta}^{(1)}|/b_{n_1} \le C < 1$ for $n$ large enough and  $\delta_{n_2} = C/\sqrt{n}$, 
$ \hat{T}^{\rm adam} \stackrel{D}{\longrightarrow} \mathcal{N}(C/\sigma_d, 1). $

{\rm (8)} Under $H_{17n}$ with $r_{n_0} \delta^4_{n_0} |\mu_{\Delta}^{(0)}|/b_{n_0} \ge C > 1$, or $r_{n_1} \delta^4_{n_1} |\mu_{\Delta}^{(1)}|/b_{n_1} \ge C > 1$ for $n$ large enough, or $\sqrt{n} \delta_{n_2} \to \infty$, $\hat{T}^{\rm adam}$ diverges to infinity in probability at least at rate $ r_{n_0} \delta_{n_0}^4 \wedge r_{n_1} \delta_{n_1}^4 \wedge \sqrt{n} \delta_{n_2}$;
under $H_{17n}$ with  $r_{n_0} \delta^4_{n_0} |\mu_{\Delta}^{(0)}|/b_{n_0} \le C < 1$, $r_{n_1} \delta^4_{n_1} |\mu_{\Delta}^{(1)}|/b_{n_1} \le C < 1$ for $n$ large enough, and $\delta_{n_2} = C/\sqrt{n}$, 
$ \hat{T}^{\rm adam} \stackrel{D}{\longrightarrow} \mathcal{N}(C/\sigma_d, 1). $
\end{theorem}
\vspace{-16pt}

\section{Application to Transfer Learning}\label{sec:applicationTL}
The proposed two-sample testing framework can be readily applied to assess the reliability of transfer learning, which aims to improve learning performance on a target task by leveraging information from one or more related source domains. Effective knowledge transfer requires sufficient structural similarity between the source and target domains, although the two domains need not be identical \citep{pan2009survey,weiss2016survey,li2022transfer,tian2023transfer}. With population $0$ as the target domain and population $1$ as the source domain, our testing framework provides a joint statistical diagnostic for two conditions: (i) correct model specification in both domains, and (ii) sufficient similarity between the two domains.

Much of the transfer learning literature characterizes or mitigates domain discrepancies through divergence metrics \citep{ben2010theory}, adversarial training \citep{ganin2016domain}, or representation alignment \citep{wang2018deep}, but implicitly assumes that the employed model is correctly specified. This assumption, however, requires verification in practice.
Ignoring model misspecification can invalidate transferability assessments. First, it threatens reliability: even when the two domains are closely related, a misspecified model (e.g., incorrect functional form or omitted variables) can yield biased parameter estimates, leading to flawed inferences in the target domain. 
Second, it threatens validity: model misspecification may artificially inflate estimated domain discrepancy, so that a test for domain similarity may reject the null due to model misfit rather than a genuine difference between the underlying regression relationships, leading to the premature exclusion of potentially informative source data. As shown by \cite{long2015learning}, transfer performance is highly sensitive to representation quality and model conditions, which underscores the need for joint diagnostics. Consequently, reliable transfer learning demands a dual validation of correct model specification and domain similarity. 

In our framework, the hypotheses $H_0^{(0)}$ and $H_0^{(1)}$ assess whether the regression models are correctly specified in the target and source domains, respectively. The functional similarity hypothesis is formulated as
$$
H_0^{(\Delta)}:
0<d\leq\Delta
\qquad\text{versus}\qquad
H_1^{(\Delta)}:
d>\Delta,
$$
where $d$ is the $L_2$-discrepancy defined in (\ref{d}) and $\Delta$ is a prespecified tolerance. Similarity therefore does not require exact equality: the two regression functions may differ, provided that their discrepancy stays within a practically acceptable range. When $d=0$, that is, when the two regression functions coincide, the similarity component in the test statistic $\hat{T}^{\rm adam}$ becomes degenerate and is not used for inference (Remark~\ref{d=0}); the restriction $0 < d$ only excludes this degenerate case.

An alternative formulation is the one-sided equivalence test 
$$
H_0^{(\Delta)}:
d\geq\Delta
\qquad\text{versus}\qquad
H_1^{(\Delta)}:
d<\Delta,
$$
in which transferability is supported only after the null hypothesis is rejected; this suits the aim of certifying similarity of the source and target domains at a prespecified level before transfer is undertaken.

Our objective is different. We view the test procedure primarily as a diagnostic for detecting discrepancy between the source and target domains. Under the widely adopted pretraining-fine-tuning paradigm, transferring information from a pretrained model is often a natural starting point for a downstream task, rather than an action taken only after exact source-target equivalence has been established. For example, pretrained language models are routinely adapted to downstream natural language processing tasks, while pretrained models in computer vision are commonly used as initialization for subsequent tasks. In such settings, the relevant inferential question is whether there is sufficient evidence to conclude that the source and target domains are too different for the intended transfer to be appropriate.
This perspective provides a natural interpretation for placing $d\leq\Delta$ in the null hypothesis. Under the null $H_0^{(\Delta)}$, the source and target regression functions are allowed to differ, but their discrepancy does not exceed the prespecified tolerance.
The alternative $H_1^{(\Delta)}$ represents a relevant departure in which the functional discrepancy is larger than the acceptable level. Thus, rather than requiring statistical evidence of similarity before transfer is
considered, the proposed test asks whether the data provide sufficient evidence of a practically meaningful source-target discrepancy that may indicate a risk of negative transfer.

The adaptive-to-model statistic developed in Section~\ref{sec:adam} can therefore be used as a pre-transfer diagnostic. Its similarity component $\hat T^{\rm ada}$ assesses whether the functional discrepancy between the source and target domains exceeds the tolerance level, while the model checking components $\hat T_0^{\rm mod}$ and $\hat T_1^{\rm mod}$ assess the model adequacy in the target and source domains, respectively.
Through the adaptive gating mechanism, the model checking components remain asymptotically inactive when the corresponding models are correctly specified and become active when sufficiently strong evidence of misspecification is present. The resulting joint procedure can therefore respond adaptively to different departures that may undermine reliable knowledge transfer.

It is important to distinguish the proposed diagnostic from a direct evaluation of the performance of a particular transfer learning algorithm. Failure to reject $H_0$ does not imply that transferring source information will necessarily improve target prediction or estimation. Rather, it indicates that the observed data provide insufficient evidence against the model-specification and functional-similarity conditions considered in the proposed framework. Conversely, rejection of $H_0$ provides evidence that at least one of these conditions is violated and therefore signals a potential risk of unreliable or negative transfer.
\vspace{-20pt}
\section{Numerical studies}

\subsection{Simulation studies}\label{sec:sim}
In this subsection, we evaluate the finite-sample performance of the proposed tests through simulations under different model settings. 
All simulation results are based on $1,000$ Monte Carlo replications at a nominal significance level of $\alpha = 0.05$. 
Across all scenarios, we use the tuning sequences $a_n = \sqrt{ \log \log (n_0 \wedge n_1)}$ and $b_{n_l} = 2 \log(n_l) $ for $l = 0, 1$. The form of $a_n$ is also used in the simulations of \cite{donald2011new}. 

To implement the proposed tests $\hat{T}^{\rm ada}$ and $\hat{T}^{\rm adam}$, we need to estimate the unknown parameter $\theta_0^{(l)}$ and the asymptotic variance $\sigma_d^2$. 
In all simulation settings, the parameters $\theta_0^{(l)}$ are estimated using the SCAD penalized likelihood estimator. Specifically, the penalized likelihood estimator $\hat{\theta}^{(l)}$ maximizes 
$$
\frac{1}{n_l}\sum_{i=1}^{n_l} J_l(X_i^{(l)}, Y_i^{(l)},\theta^{(l)}) -  \sum_{j=1}^p \rho_{\lambda_n}(|\theta_j^{(l)}|),
$$ 
where $\rho_{\lambda_n}$ denotes the SCAD penalty.
In the following simulation studies, the criterion function is taken to be $J_l(X,Y,\theta) \propto -\frac{1}{2}(Y-g_l(X^\top\theta))^2$, where $g_l(t)=t$ for Gaussian linear models considered in Studies 1 and 2, and $g_l(t)=\exp(t)$ in Study 3.

We estimate the asymptotic variance \(\sigma_d^2\) by a plug-in estimator. First, we describe the variance decomposition used for its numerical implementation. Define
$$
G_0=\zeta M_{00}+(1-\zeta)M_{10},
\qquad
G_1=\zeta M_{01}+(1-\zeta)M_{11},
$$
where $\{ M_{kl}: k,l = 0,1\}$ are the quantities defined in Theorem~\ref{th:dhatasymnormal}. 
Under the model-specification null $H_0^{(l)}$, we write
$$
Y^{(l)} =
g_l((X^{(l)})^\top\theta_0^{(l)}) + \epsilon^{(l)},
\quad \text{and} \quad
E[\epsilon^{(l)} | X^{(l)}] = 0.
$$
For the criterion function $J_l$, the score evaluated at the true parameter is
$$
\nabla_{\theta^{(l)}}J_l
(X^{(l)},Y^{(l)},\theta_0^{(l)})
=
X^{(l)}
\dot g_l\left((X^{(l)})^\top\theta_0^{(l)}\right)
\epsilon^{(l)}.
$$
By the oracle property of the SCAD estimator under suitable regularity conditions, on the true active set
$S^{(l)}$ the function $L_l(X^{(l)},Y^{(l)},\theta_0^{(l)})$ in (\ref{theta-expansion}) has the score-type representation
\[
L_{l,S^{(l)}}(X^{(l)},Y^{(l)},\theta_0^{(l)})
=
I_l^{-1}
X_{S^{(l)}}^{(l)}
\dot g_l\left((X^{(l)})^\top\theta_0^{(l)}\right)
\epsilon^{(l)},
\]
with zero entries on the inactive coordinates, where $I_l$ denotes the
corresponding information matrix restricted to $S^{(l)}$. Consequently,
$E[L_l(X^{(l)},Y^{(l)},\theta_0^{(l)})\mid X^{(l)}]=0.$
Recall that $\psi(x,\theta^{(0)},\theta^{(1)})
=g_0(x^\top\theta^{(0)})-g_1(x^\top\theta^{(1)}).$
It follows that 
$$
\operatorname{Cov}\left[
\psi^2(X^{(l)},\theta_0^{(0)},\theta_0^{(1)}),
G_l^\top
L_l(X^{(l)},Y^{(l)},\theta_0^{(l)})
\right]
=0,
\qquad l=0,1.
$$
Therefore, under the joint model-specification null
$H_0^{(0)}\cap H_0^{(1)}$, the asymptotic variance $\sigma_d^2$ defined in Theorem~\ref{th:dhatasymnormal} admits the simplified decomposition
\begin{align}
\sigma_d^2
=& \zeta \operatorname{Var}\left[\psi^2(X^{(0)}, \theta_0^{(0)}, \theta_0^{(1)})\right]+
   (1-\zeta)\operatorname{Var}\left[\psi^2(X^{(1)}, \theta_0^{(0)}, \theta_0^{(1)})\right] \nonumber \\
&  +\frac{4}{\zeta}G_0^\top\Sigma_0G_0 + \frac{4}{1-\zeta}G_1^\top\Sigma_1G_1,
\label{eq:sigmad}
\end{align}
where $ \Sigma_l = E[ L_l(X^{(l)},Y^{(l)},\theta_0^{(l)}) L_l(X^{(l)},Y^{(l)},\theta_0^{(l)})^\top] $.

We next construct the estimator of each component of $\sigma^2_d$ in (\ref{eq:sigmad}).
Based on the oracle property of the SCAD estimator, the function $L_l(X^{(l)},Y^{(l)},\theta_0^{(l)})$ is
estimated on the selected active set $\hat S^{(l)}$ as
\[
\hat L_l(X_i^{(l)},Y_i^{(l)},\hat\theta^{(l)})
=
\begin{pmatrix}
\hat I^{-1}_{l}
\nabla_{\theta_{\hat S^{(l)}}}
J_l
(X_i^{(l)},Y_i^{(l)},\hat\theta_{\hat S^{(l)}}^{(l)})
\\
\mathbf{0}_{p-|\hat S^{(l)}|}
\end{pmatrix},
\]
where
\(
\hat I_l
=
-n_l^{-1}
\sum_{i=1}^{n_l}
\nabla^2_{\theta_{\hat S^{(l)}}}
J_l
(X_i^{(l)},Y_i^{(l)},\hat\theta_{\hat S^{(l)}}^{(l)})
\)
is the empirical information matrix restricted to the selected active
set and $ \nabla_{\theta_{\hat S^{(l)}}} J_l (X_i^{(l)},Y_i^{(l)},\hat\theta_{\hat S^{(l)}}^{(l)}) $ denotes the gradient of $J_l (X_i^{(l)},Y_i^{(l)},\hat\theta_{\hat S^{(l)}}^{(l)})$ with respect to $\theta_{\hat{S}^{(l)}}$.
Thus, the covariance matrix $\Sigma_l$ is estimated by
$$
\hat\Sigma_l=
\frac{1}{n_l}\sum_{i=1}^{n_l}
\hat L_l(X_i^{(l)},Y_i^{(l)},\hat\theta^{(l)})
\hat L_l(X_i^{(l)},Y_i^{(l)},\hat\theta^{(l)})^\top, \quad l=0, 1.
$$
Define
$
\hat\psi(x)
=\psi(x,\hat\theta^{(0)},\hat\theta^{(1)})
=g_0(x^\top\hat\theta^{(0)})-g_1(x^\top\hat\theta^{(1)}).
$
The plug-in estimator of $\sigma_d^2$ is obtained by replacing all
population quantities in~(\ref{eq:sigmad}) with their sample estimators,
\begin{align}
\hat\sigma_d^2
=
\frac{n_0}{n}
\widehat{\operatorname{Var}}
\left[\hat\psi^2(X^{(0)})\right]
+
\frac{n_1}{n}
\widehat{\operatorname{Var}}
\left[\hat\psi^2(X^{(1)})\right]
+
\frac{4n}{n_0}
\hat G_0^\top\hat\Sigma_0\hat G_0
+
\frac{4n}{n_1}
\hat G_1^\top\hat\Sigma_1\hat G_1 ,
\end{align}
where $\hat G_0$ and $\hat G_1$ are the corresponding empirical analogues of
$G_0$ and $G_1$, respectively. This plug-in estimator is used in the construction of
$\hat T^{\rm LFC}$, $\hat T^{\rm ada}$, and $\hat{T}^{\rm adam}$ throughout all simulation studies. Under correct model specification, this plug-in estimator is consistent for the asymptotic variance, since the cross-covariance term in Theorem~\ref{th:dhatasymnormal} vanishes. Under model misspecification in Studies 2 and 3, it serves as an approximation, with rejection primarily driven by the activated model-specification component rather than the similarity component.

We first investigate the performance of the adaptive boundary-proxy test $\hat{T}^{\rm ada}$ for assessing functional similarity between the two populations. We also compare our test $\hat{T}^{\rm ada}$ with the classical LFC-based test $\hat{T}^{\rm LFC}$.

{\it Study 1.} The data are generated from the following Gaussian linear models for the two populations
\begin{align}\label{study1}
\begin{split}
    Y^{(0)} &= (X^{(0)})^\top \theta_{0}^{(0)} + \epsilon^{(0)}, \\
    Y^{(1)} &= (X^{(1)})^\top \theta_{0}^{(1)} + \epsilon^{(1)},
\end{split}
\end{align}
with $X^{(l)} \sim \mathcal{N}(0, \mathbf{I}_p)$ and $\epsilon^{(l)} \sim \mathcal{N}(0, 1)$. The regression coefficient vectors for populations $0$ and $1$ are $\theta_{0}^{(0)} = (1, 1, 1, 1, 1, 0, \dots, 0)^\top$ and $\theta_{0}^{(1)} = (1.5, 1.5, 1.5, 1.5, 1.5, 0, \dots, 0)^\top$, respectively. Elementary calculations show that $d = \int_\mathcal{X} |g_0(x^{\top} \theta_0^{(0)})-g_1(x^{\top} \theta_0^{(1)})|^2 dF(x) = \| \theta_0^{(0)} - \theta_0^{(1)} \|_2^2 = 1.25$. 
The predictor dimensions are $p \in \{50, 200, 600\}$ and the sample sizes are set to $n_0 \in \{200, 300\}$ and $n_1 \in \{500, 800\}$.

\begin{table}[htbp]
\centering
\small
\caption{Empirical size and power of the proposed and LFC tests in Study 1 ($d=1.25$).}
\label{tab:study1}
\resizebox{\textwidth}{!}{
\begin{tabular}{cccc cccc cccc}
\toprule
\multirow{2}{*}{$p$}
& \multirow{2}{*}{$n_1$}
& \multirow{2}{*}{$n_0$}
& \multirow{2}{*}{Method}
& \multicolumn{4}{c}{Empirical size}
& \multicolumn{4}{c}{Empirical power} \\
\cmidrule(lr){5-8}\cmidrule(lr){9-12}

&
&
&
& $\Delta=3$
& $\Delta=1.5$
& $\Delta=1.3$
& $\Delta=d$
& $\Delta=1.2$
& $\Delta=1.1$
& $\Delta=1.0$
& $\Delta=0.8$ \\
\midrule

\multirow{8}{*}{50}
& \multirow{2}{*}{500}
& \multirow{2}{*}{200}
& $\hat{T}^{\rm ada}$ 
& 0.044 & 0.022 & 0.041 & 0.061
& 0.095 & 0.207 & 0.395 & 0.825 \\
&
&
& $\hat{T}^{\rm LFC}$      
& 0.000 & 0.001 & 0.033 & 0.057
& 0.093 & 0.207 & 0.395 & 0.825 \\

& \multirow{2}{*}{500}
& \multirow{2}{*}{300}
& $\hat{T}^{\rm ada}$ 
& 0.047 & 0.020 & 0.028 & 0.054
& 0.085 & 0.224 & 0.459 & 0.900 \\
&
&
& $\hat{T}^{\rm LFC}$      
& 0.000 & 0.001 & 0.022 & 0.049
& 0.084 & 0.224 & 0.459 & 0.900 \\

& \multirow{2}{*}{800}
& \multirow{2}{*}{200}
& $\hat{T}^{\rm ada}$ 
& 0.057 & 0.029 & 0.035 & 0.053
& 0.090 & 0.217 & 0.423 & 0.850 \\
&
&
& $\hat{T}^{\rm LFC}$      
& 0.000 & 0.002 & 0.028 & 0.049
& 0.087 & 0.217 & 0.423 & 0.850 \\

& \multirow{2}{*}{800}
& \multirow{2}{*}{300}
& $\hat{T}^{\rm ada}$ 
& 0.053 & 0.020 & 0.035 & 0.063
& 0.113 & 0.282 & 0.532 & 0.938 \\
&
&
& $\hat{T}^{\rm LFC}$      
& 0.000 & 0.000 & 0.029 & 0.060
& 0.112 & 0.282 & 0.532 & 0.938 \\[1mm]

\midrule

\multirow{8}{*}{200}
& \multirow{2}{*}{500}
& \multirow{2}{*}{200}
& $\hat{T}^{\rm ada}$ 
& 0.061 & 0.030 & 0.039 & 0.058
& 0.089 & 0.220 & 0.391 & 0.834 \\
&
&
& $\hat{T}^{\rm LFC}$      
& 0.000 & 0.003 & 0.032 & 0.054
& 0.088 & 0.220 & 0.391 & 0.834 \\

& \multirow{2}{*}{500}
& \multirow{2}{*}{300}
& $\hat{T}^{\rm ada}$ 
& 0.045 & 0.026 & 0.038 & 0.062
& 0.103 & 0.254 & 0.465 & 0.893 \\
&
&
& $\hat{T}^{\rm LFC}$      
& 0.000 & 0.000 & 0.028 & 0.056
& 0.099 & 0.253 & 0.465 & 0.893 \\

& \multirow{2}{*}{800}
& \multirow{2}{*}{200}
& $\hat{T}^{\rm ada}$ 
& 0.054 & 0.025 & 0.045 & 0.073
& 0.117 & 0.245 & 0.438 & 0.880 \\
&
&
& $\hat{T}^{\rm LFC}$      
& 0.000 & 0.003 & 0.037 & 0.069
& 0.114 & 0.245 & 0.438 & 0.880 \\

& \multirow{2}{*}{800}
& \multirow{2}{*}{300}
& $\hat{T}^{\rm ada}$ 
& 0.043 & 0.021 & 0.048 & 0.071
& 0.116 & 0.272 & 0.535 & 0.932 \\
&
&
& $\hat{T}^{\rm LFC}$      
& 0.000 & 0.001 & 0.040 & 0.065
& 0.113 & 0.272 & 0.535 & 0.932 \\[1mm]

\midrule

\multirow{8}{*}{600}
& \multirow{2}{*}{500}
& \multirow{2}{*}{200}
& $\hat{T}^{\rm ada}$ 
& 0.056 & 0.025 & 0.051 & 0.065
& 0.098 & 0.215 & 0.412 & 0.847 \\
&
&
& $\hat{T}^{\rm LFC}$      
& 0.000 & 0.003 & 0.042 & 0.060
& 0.095 & 0.214 & 0.412 & 0.847 \\

& \multirow{2}{*}{500}
& \multirow{2}{*}{300}
& $\hat{T}^{\rm ada}$ 
& 0.061 & 0.027 & 0.033 & 0.061
& 0.094 & 0.241 & 0.492 & 0.909 \\
&
&
& $\hat{T}^{\rm LFC}$      
& 0.000 & 0.000 & 0.028 & 0.059
& 0.092 & 0.241 & 0.492 & 0.909 \\

& \multirow{2}{*}{800}
& \multirow{2}{*}{200}
& $\hat{T}^{\rm ada}$ 
& 0.064 & 0.021 & 0.045 & 0.065
& 0.104 & 0.254 & 0.454 & 0.881 \\
&
&
& $\hat{T}^{\rm LFC}$      
& 0.000 & 0.001 & 0.041 & 0.063
& 0.102 & 0.253 & 0.454 & 0.881 \\

& \multirow{2}{*}{800}
& \multirow{2}{*}{300}
& $\hat{T}^{\rm ada}$ 
& 0.068 & 0.036 & 0.044 & 0.068
& 0.112 & 0.263 & 0.517 & 0.934 \\
&
&
& $\hat{T}^{\rm LFC}$      
& 0.000 & 0.001 & 0.036 & 0.065
& 0.110 & 0.262 & 0.517 & 0.934 \\

\bottomrule
\end{tabular}
}
\end{table}

Table~\ref{tab:study1} compares the finite-sample performance of our adaptive test $\hat{T}^{\rm ada}$  and the classic LFC-based test $\hat{T}^{\rm LFC}$ with $d=1.25$.  
First, under the null $H_0^{(\Delta)}$, at the boundary $\Delta = 1.25$, the proposed test $\hat{T}^{\rm ada}$ exhibits a slightly higher rejection rate than the LFC-based test, while both remain close to the nominal level $\alpha$ across all configurations.
When $\Delta$ is only slightly larger than $d$ ($\Delta = 1.3,1.5$), the rejection rate of the proposed test $\hat{T}^{\rm ada}$ decreases but remains higher than that of the LFC test. When $\Delta$ is substantially larger than $d$ ($\Delta = 3$), the LFC test's rejection rate quickly drops to zero, reflecting its increasingly conservative nature. In contrast, at $\Delta=3$, the rejection rate of the proposed test remains close to the nominal level of 0.05.
Second, under the alternative $(\Delta = 0.8, 1.0, 1.1,1.2)$, the proposed test and the LFC test yield nearly identical rejection rates across all settings, and the power increases with larger sample sizes. This agreement is expected because, with $a_n = \sqrt{\log \log (n_0 \wedge n_1)}$, the adaptive statistic tends to select the LFC component by design under the alternative hypothesis.
Moreover, the results remain stable as the dimension increases from $p=50$ to $p=600$, indicating satisfactory performance in high-dimensional settings.

After demonstrating the adaptive property of our test through comparison with the LFC benchmark, we now systematically evaluate its performance in two-sample regression settings under distinct scenarios. These scenarios are designed to challenge different aspects of the testing procedure, including high dimensionality, feature correlation, and different model misspecification modes. 

{\it Study 2.} The data are generated from the following regression
models for the two populations
\begin{align}
\begin{split}
    Y^{(0)} &= (X^{(0)})^\top \theta_{0}^{(0)} + w_{0}(X^{(0)}) + \epsilon^{(0)}, \\
    Y^{(1)} &= (X^{(1)})^\top \theta_{0}^{(1)} + w_{1}(X^{(1)}) + \epsilon^{(1)},
\end{split}
\end{align}
with $\epsilon^{(l)} \sim \mathcal{N}(0, 1)$ and $X^{(l)} \sim \mathcal{N}(0, \mathbf{\Sigma})$. The function $w_l(\cdot)$ captures potential model misspecification.
The regression coefficient vectors for populations $0$ and $1$ are $\theta_{0}^{(0)} = (1, 1, 1, 1, 1, 0, \dots, 0)^\top$ and $\theta_{0}^{(1)} = (1.5, 1.5, 1.5, 1.5, 1.5, 0, \dots, 0)^\top$, respectively. 
The predictor dimensions $p \in \{50, 200, 600\}$ and the sample sizes are the same as in Study~1.
For each setting, we consider two nonlinear model misspecification terms:
\begin{align*}
    w_l(X) &= c \cdot \cos(0.6\pi X^\top \theta_{0}^{(l)}),\\
    w_l(X) &= 0.1c \cdot (X^\top \theta_{0}^{(l)})^2,
\end{align*}
where $c \in \{0, 0.5, 1\}$ controls the magnitude of nonlinearity. The case $c=0$ corresponds to the correctly specified linear model.
Moreover, we consider two covariance structures for $X^{(l)}$, $ \mathbf{\Sigma}=\mathbf{I}_p$ and
$\mathbf{\Sigma} = (\rho^{|i-j|})_{p\times p}$ with $\rho=0.4$. Under these two covariance structures, the corresponding true discrepancies under the joint null $H_0^{(0)} \cap H_0^{(1)}$ are $d=1.25$ and $d\approx 2.3668$, respectively. 

\begin{table}[htbp]
\centering
\small
\caption{Empirical size and power under independent covariates in Study 2
($d=1.25$).}
\label{tab:study2_independent}
\resizebox{\textwidth}{!}{
\begin{tabular}{ccc cccc cc cc cc}
\toprule
\multirow{4}{*}{$p$}
& \multirow{4}{*}{$n_1$}
& \multirow{4}{*}{$n_0$}
& \multicolumn{4}{c}{\multirow{2}{*}{Empirical size}}
& \multicolumn{2}{c}{\multirow{2}{*}{Similarity power}}
& \multicolumn{4}{c}{Model misspecification power} \\
\cmidrule(lr){10-13}

&
&
& \multicolumn{4}{c}{}
& \multicolumn{2}{c}{}
& \multicolumn{2}{c}{Cosine}
& \multicolumn{2}{c}{Quadratic} \\
\cmidrule(lr){4-7}
\cmidrule(lr){8-9}
\cmidrule(lr){10-11}
\cmidrule(lr){12-13}

&
&
& $\Delta=3$
& $\Delta=1.5$
& $\Delta=1.3$
& $\Delta=d$
& $\Delta=1.2$
& $\Delta=1.0$
& $\Delta=3$
& $\Delta=3$
& $\Delta=3$
& $\Delta=3$ \\

&
&
& $c=0$ & $c=0$ & $c=0$ & $c=0$ & $c=0$ & $c=0$ & $c=0.5$
& $c=1$
& $c=0.5$
& $c=1$ \\
\midrule

\multirow{4}{*}{50}
& 500 & 200
& 0.047 & 0.028 & 0.042 & 0.065
& 0.089 & 0.399
& 0.420 & 0.997
& 0.276 & 0.995 \\
& 500 & 300
& 0.044 & 0.027 & 0.038 & 0.049
& 0.100 & 0.467
& 0.424 & 0.996
& 0.740 & 1.000 \\
& 800 & 200
& 0.055 & 0.033 & 0.050 & 0.078
& 0.108 & 0.434
& 0.936 & 1.000
& 0.455 & 1.000 \\
& 800 & 300
& 0.055 & 0.031 & 0.038 & 0.068
& 0.100 & 0.525
& 0.924 & 1.000
& 0.783 & 1.000 \\[1mm]

\multirow{4}{*}{200}
& 500 & 200
& 0.063 & 0.025 & 0.045 & 0.072
& 0.084 & 0.425
& 0.170 & 0.793
& 0.114 & 0.840 \\
& 500 & 300
& 0.061 & 0.027 & 0.037 & 0.064
& 0.093 & 0.445
& 0.184 & 0.813
& 0.387 & 0.978 \\
& 800 & 200
& 0.053 & 0.027 & 0.027 & 0.053
& 0.105 & 0.433
& 0.752 & 0.999
& 0.273 & 1.000 \\
& 800 & 300
& 0.062 & 0.028 & 0.030 & 0.058
& 0.097 & 0.523
& 0.747 & 1.000
& 0.480 & 1.000 \\[1mm]

\multirow{4}{*}{600}
& 500 & 200
& 0.055 & 0.038 & 0.044 & 0.073
& 0.112 & 0.427
& 0.116 & 0.360
& 0.069 & 0.591 \\
& 500 & 300
& 0.045 & 0.022 & 0.035 & 0.055
& 0.092 & 0.489
& 0.096 & 0.375
& 0.227 & 0.806 \\
& 800 & 200
& 0.063 & 0.028 & 0.040 & 0.086
& 0.118 & 0.461
& 0.428 & 0.951
& 0.190 & 0.977 \\
& 800 & 300
& 0.052 & 0.026 & 0.036 & 0.070
& 0.105 & 0.507
& 0.428 & 0.964
& 0.300 & 0.993 \\

\bottomrule
\end{tabular}
}
\end{table}

\begin{table}[htbp]
\centering
\small
\caption{Empirical size and power under correlated covariates in Study 2
($d\approx 2.37$).}
\label{tab:study2_correlated}
\resizebox{\textwidth}{!}{
\begin{tabular}{ccc cccc cc cc cc}
\toprule
\multirow{4}{*}{$p$}
& \multirow{4}{*}{$n_1$}
& \multirow{4}{*}{$n_0$}
& \multicolumn{4}{c}{\multirow{2}{*}{Empirical size}}
& \multicolumn{2}{c}{\multirow{2}{*}{Similarity power}}
& \multicolumn{4}{c}{Model misspecification power} \\
\cmidrule(lr){10-13}

&
&
& \multicolumn{4}{c}{}
& \multicolumn{2}{c}{}
& \multicolumn{2}{c}{Cosine}
& \multicolumn{2}{c}{Quadratic} \\
\cmidrule(lr){4-7}
\cmidrule(lr){8-9}
\cmidrule(lr){10-11}
\cmidrule(lr){12-13}

&
&
& $\Delta=5$
& $\Delta=3$
& $\Delta=2.5$
& $\Delta=d$
& $\Delta=2$
& $\Delta=1.5$
& $\Delta=5$
& $\Delta=5$
& $\Delta=5$
& $\Delta=5$ \\

&
&
& $c=0$
& $c=0$
& $c=0$
& $c=0$
& $c=0$
& $c=0$
& $c=0.5$
& $c=1$
& $c=0.5$
& $c=1$ \\
\midrule

\multirow{4}{*}{50}
& 500 & 200
& 0.042 & 0.044 & 0.034 & 0.051
& 0.381 & 0.965
& 0.379 & 0.997
& 0.989 & 1.000 \\
& 500 & 300
& 0.046 & 0.045 & 0.026 & 0.050
& 0.432 & 0.991
& 0.341 & 0.999
& 1.000 & 1.000 \\
& 800 & 200
& 0.056 & 0.045 & 0.034 & 0.075
& 0.417 & 0.979
& 0.888 & 1.000
& 1.000 & 1.000 \\
& 800 & 300
& 0.061 & 0.043 & 0.017 & 0.058
& 0.508 & 0.992
& 0.879 & 1.000
& 1.000 & 1.000 \\[1mm]

\multirow{4}{*}{200}
& 500 & 200
& 0.055 & 0.038 & 0.031 & 0.073
& 0.397 & 0.956
& 0.236 & 0.944
& 0.876 & 1.000 \\
& 500 & 300
& 0.048 & 0.045 & 0.021 & 0.051
& 0.446 & 0.987
& 0.229 & 0.930
& 0.985 & 1.000 \\
& 800 & 200
& 0.046 & 0.042 & 0.027 & 0.057
& 0.429 & 0.981
& 0.757 & 1.000
& 0.999 & 1.000 \\
& 800 & 300
& 0.072 & 0.042 & 0.020 & 0.065
& 0.502 & 0.997
& 0.766 & 1.000
& 1.000 & 1.000 \\[1mm]

\multirow{4}{*}{600}
& 500 & 200
& 0.051 & 0.040 & 0.032 & 0.059
& 0.411 & 0.969
& 0.145 & 0.708
& 0.756 & 0.996 \\
& 500 & 300
& 0.062 & 0.047 & 0.025 & 0.054
& 0.464 & 0.978
& 0.132 & 0.728
& 0.925 & 0.999 \\
& 800 & 200
& 0.045 & 0.041 & 0.030 & 0.055
& 0.435 & 0.979
& 0.613 & 1.000
& 0.984 & 1.000 \\
& 800 & 300
& 0.047 & 0.049 & 0.029 & 0.059
& 0.501 & 0.995
& 0.579 & 0.998
& 0.997 & 1.000 \\

\bottomrule
\end{tabular}
}
\end{table}

Table \ref{tab:study2_independent} and Table \ref{tab:study2_correlated} show the empirical size and power of our proposed test $\hat{T}^{\rm adam}$ under independent and correlated covariate structures, respectively. Under the correctly specified models, the empirical rejection rates are generally well controlled across different values of dimension and sample sizes. 
At the boundary \(\Delta=d\), the rejection rates are generally close to the nominal level, although some configurations exhibit modest size inflation. When \(\Delta\) is only slightly larger than \(d\), conservative rejection rates are observed. As \(\Delta-d\) increases further, the rejection rates return closer to the nominal level. This pattern is consistent with the adaptive switching mechanism.
Under the similarity alternatives with $\Delta<d$, as expected, power generally increases with larger sample sizes and a wider gap between $d$ and $\Delta$. 
Under the misspecification alternatives, power against both cosine and quadratic misspecification generally increases with $c$ and sample size, although it remains limited in some weak-signal, high-dimensional settings.
Overall, the results demonstrate that the test can achieve satisfactory size control and power against both distance violations and model misspecification.

Next, we consider a more challenging setting, where the conditional mean function follows a generalized linear model with an exponential link.

{\it Study 3.} The data are generated from the following regression
models for the two populations
\begin{align}
\begin{split}
    Y^{(0)}|X^{(0)} &\sim \mathcal{N}\bigl(\exp\bigl((X^{(0)})^\top \theta_{0}^{(0)} + \tilde{w}_{0}(X^{(0)})\bigr), \sigma^2\bigr), \\
    Y^{(1)}|X^{(1)} &\sim \mathcal{N}\bigl(\exp\bigl((X^{(1)})^\top \theta_{0}^{(1)} + \tilde{w}_{1}(X^{(1)})\bigr), \sigma^2\bigr),
\end{split}
\end{align}
where $X^{(l)} \sim \mathcal{N}(0, \mathbf{I}_p)$ and the standard deviation $\sigma = 0.5$ is set to ensure numerical stability. The regression coefficient vectors for populations $0$ and $1$ are $\theta_{0}^{(0)} = (0.1, 0.1, 0.1, 0.1, 0.1, 0, \dots, 0)^\top$ and $\theta_{0}^{(1)} = (0.4, 0.4, 0.4, 0.4, 0.4, 0, \dots, 0)^\top$, respectively. The misspecification term $\tilde{w}_l(X) = 0.4c \cdot \cos(0.6\pi X^\top \theta_{0}^{(l)})$ is added inside the exponential function, with $c \in \{0, 0.5, 1\}$.
Under the correct model specification, the true functional discrepancy is analytically computed as $d \approx 2.3217$.
The predictor dimensions $p \in \{50, 200, 600\}$ and the sample sizes are set to $n_0 \in \{200, 300\}$ and $n_1 \in \{500, 800\}$. Tolerance thresholds are set to $\Delta \in \{d,2.5, 3, 5\}$ for size since $d \approx 2.32$, $\Delta \in \{1.5, 2\}$ for power against distance violation, and $\Delta = 5$ with $c \in \{0.5, 1\}$ for power against misspecification.

\begin{table}[htbp]
\centering
\small
\caption{Empirical size and power in Study 3 ($d\approx2.32$).}
\label{tab:study3}
\resizebox{\textwidth}{!}{
\begin{tabular}{ccc cccc cccc}
\toprule
\multirow{3}{*}{$p$}
& \multirow{3}{*}{$n_1$}
& \multirow{3}{*}{$n_0$}
& \multicolumn{4}{c}{Empirical size}
& \multicolumn{4}{c}{Empirical power} \\
\cmidrule(lr){4-7}\cmidrule(lr){8-11}

&
&
& $\Delta=5$ & $\Delta=3$ & $\Delta=2.5$ & $\Delta=d$
& $\Delta=2$ & $\Delta=1.5$ & $\Delta=5$ & $\Delta=5$ \\

&
&
& $c=0$ & $c=0$ & $c=0$ & $c=0$
& $c=0$ & $c=0$ & $c=0.5$ & $c=1$ \\
\midrule

\multirow{4}{*}{50}
& 500 & 200
& 0.056 & 0.030 & 0.033 & 0.034
& 0.157 & 0.641 & 0.162 & 0.981 \\
& 500 & 300
& 0.061 & 0.043 & 0.036 & 0.033
& 0.104 & 0.603 & 0.155 & 0.977 \\
& 800 & 200
& 0.047 & 0.039 & 0.039 & 0.048
& 0.237 & 0.801 & 0.385 & 1.000 \\
& 800 & 300
& 0.055 & 0.029 & 0.028 & 0.036
& 0.140 & 0.707 & 0.378 & 1.000 \\[1mm]

\multirow{4}{*}{200}
& 500 & 200
& 0.079 & 0.050 & 0.047 & 0.061
& 0.182 & 0.722 & 0.169 & 0.872 \\
& 500 & 300
& 0.057 & 0.037 & 0.026 & 0.039
& 0.152 & 0.653 & 0.149 & 0.839 \\
& 800 & 200
& 0.082 & 0.050 & 0.054 & 0.092
& 0.303 & 0.865 & 0.255 & 1.000 \\
& 800 & 300
& 0.061 & 0.040 & 0.026 & 0.048
& 0.215 & 0.793 & 0.266 & 0.997 \\[1mm]

\multirow{4}{*}{600}
& 500 & 200
& 0.053 & 0.043 & 0.047 & 0.072
& 0.232 & 0.805 & 0.175 & 0.732 \\
& 500 & 300
& 0.064 & 0.032 & 0.035 & 0.044
& 0.169 & 0.713 & 0.158 & 0.724 \\
& 800 & 200
& 0.057 & 0.039 & 0.063 & 0.087
& 0.358 & 0.907 & 0.291 & 0.988 \\
& 800 & 300
& 0.066 & 0.032 & 0.035 & 0.072
& 0.238 & 0.824 & 0.333 & 0.988 \\

\bottomrule
\end{tabular}
}
\end{table}

The results from Table~\ref{tab:study3} show that the proposed test exhibits good power against both distance violations and model misspecification under the challenging exponential GLM setting.
At the boundary $\Delta=d$, we observe a moderate size inflation in some high-dimensional settings. This phenomenon may reflect the finite-sample bias in the parameter estimates under the exponential link.

\subsection{Real Data Analysis}
In this subsection, we illustrate the application of the proposed two-sample testing framework to transfer learning using a real-world housing price dataset. While the preceding simulations evaluate the finite-sample performance of the proposed test under controlled data-generating processes, this analysis demonstrates its practical utility in assessing model specification and source-target functional similarity when the underlying regression relationships and their discrepancy are unknown.
We use the Ames Housing dataset, which records residential property sales in Ames, Iowa from 2006 to 2010. It includes detailed information on housing characteristics relevant to property value assessment. The data are available at \url{https://www.kaggle.com/competitions/house-prices-advanced-regression-techniques}.
The response variable is the house sale price, and we use its log-transformed value as the response in the regression model. The original dataset contains 79 explanatory variables describing property characteristics, including lot size, neighborhood, overall quality, construction year, living area, garage information, sale condition and other housing-related attributes. After preprocessing categorical variables through dummy encoding, the resulting predictor dimension is $p=261$. 

To construct a natural transfer learning setting, we split the data according to the year of sale. Houses sold during 2006-2008 are used as the source domain, while houses sold during 2009-2010 are used as the target domain. This setup represents a transfer problem: we aim to examine whether the housing price regression relationship learned from an earlier market period can be reliably transferred to a subsequent market period. This division gives source and target sample sizes of $n_1=947$ and $n_0=513$, respectively.

We specify linear regression models for both the source and target domains. In practice, $\Delta$ represents a substantive tolerance threshold, which ideally should be determined by expert or regulatory knowledge. In the absence of such external guidance, we address the scale of the response by specifying the similarity tolerance relative to the empirical variance of the log-transformed sale price. Specifically, we set $\Delta = \rho \cdot \widehat{\operatorname{var}}(Y)$, where $\rho$ is a proportion parameter and $\widehat{\operatorname{var}}(Y)$ is the sample variance of the response $Y$. This makes the tolerance level easier to interpret and avoids relying on an arbitrary fixed constant. In the following analysis, we consider four pre-specified tolerance levels $\rho \in \{0.02,0.05, 0.10, 0.20\}$, which range from strict to liberal transferability criteria.

\begin{table}[htbp]
\centering
\caption{Test on the Ames Housing dataset under different tolerance levels. The last two columns are computed from one fixed sample split; see the text.}
\label{tab:realdata_ames}
\begin{tabular}{cccccc}
\hline
$\rho$ & $\Delta$ & $\hat d$ & $\hat T^{\rm{LFC}}$ 
& $\hat T^{\rm{adam}}$ & Decision \\
\hline
$0.02$  & $0.0032$ & $0.0105$ & $1.9255$  & $1.9255$  & Reject \\
$0.05$  & $0.0080$ & $0.0105$ & $0.6710$  & $0.6710$  & Not reject \\
$0.10$ & $0.0160$ & $0.0105$ & $-1.4197$ & $-3.3989$ & Not reject \\
$0.20$ & $0.0319$ & $0.0105$ & $-5.6013$ & $-3.3989$ & Not reject \\
\hline
\end{tabular}
\end{table}

The estimated linear regression function distance between the source and target domains is $\hat d=0.0105$. The tuning parameter for the adaptive similarity test is $a_n=1.3532$, with the corresponding switching threshold $-a_n=-1.3532$, and the Error-Proxy statistic is $\hat T_c=-3.3989$. The model-specification statistics are $\hat T^{\rm mod}_0=2.0201$ and $\hat T^{\rm mod}_1=3.3375$. Both statistics fall below their corresponding truncation thresholds, $b_{n_0}=12.4806$ and $b_{n_1}=13.7066$, respectively. Therefore, the model misspecification terms are not activated.
As Table~\ref{tab:realdata_ames} shows, the proposed test rejects the transferability null only under the most stringent tolerance level, $\rho=2\%$. For all tolerance levels $\rho\geq 5\%$, the adaptive test statistic $\hat T^{\rm adam}$ remains below the critical value of $1.6449$, and hence the null hypothesis is not rejected. This result suggests that the source and target housing price models are not adequately similar under an extremely strict similarity requirement. In contrast, under moderate and practically meaningful tolerance levels, the data do not provide sufficient evidence that the functional discrepancy exceeds the specified threshold.

In summary, the Ames Housing analysis demonstrates the practical interpretation of the proposed test. Rather than testing for exact equality, the proposed procedure evaluates whether the discrepancy exceeds a user-specified tolerance level. In other words, it separates differences that are practically meaningful from those that are negligible relative to the specified margin. As noted in Section~\ref{sec:applicationTL}, however, non-rejection does not guarantee that transfer is safe.

\section{Discussion}
The methodology developed in this paper is formulated for the comparison of two populations. Several extensions are worth further investigation.

First, the framework could be extended beyond parametric regression models to more flexible semiparametric and nonparametric settings, such as regression functions represented in a reproducing kernel Hilbert space. Such extensions would require new tools for estimating the regression function discrepancy and deriving its asymptotic distribution.
Second, the automated selection of the similarity threshold $\Delta$ represents a critical practical challenge. While we adopted a variance-proportional approach in our empirical analysis, a purely data-driven method to determine the optimal $\Delta$ remains elusive. 
In the transfer learning setting, one possible direction is to explore the trade-off between the risk of negative transfer and the potential gain from source knowledge to provide a more objective criterion for threshold setting.
Finally, the transfer learning application discussed in Section~\ref{sec:applicationTL} naturally motivates an extension from a single source domain to multiple source domains.
Although transfer learning provides an important application of the proposed framework, the testing methodology itself is not restricted to this setting. The same formulation can be used whenever the objective is to assess whether two regression relationships are sufficiently close while simultaneously assessing the specification of the corresponding models. Exploring other two-sample regression problems may therefore provide additional applications of the proposed methodology.
\vspace{-12pt}

\bibliographystyle{abbrvnat}
\bibliography{ref}

\end{document}